\documentclass[aps, prb, amsmath, amssymb, reprint, longbibliography]{revtex4-2}
\usepackage{graphicx}
\usepackage{dcolumn}
\usepackage{bm}
\usepackage{color}
\usepackage{braket}
\usepackage{multirow}
\usepackage{amsmath,amssymb,amsfonts,bm}

\begin{document}

\title{Dielectric screening at TMD:hBN interfaces: Monolayer-to-bulk transition, local-field effect, and spatial dependence}
\author{Olugbenga Adeniran}
\author{Zhen-Fei Liu}
\email{zfliu@wayne.edu}
\affiliation{Department of Chemistry, Wayne State University, Detroit, Michigan 48202 USA}

\date{\today}

\begin{abstract}
The dielectric effects of a substrate have been shown to be important in modulating the electronic properties of an adsorbate, especially in van der Waals heterostructures. Here, using the first-principles dielectric embedding $GW$ approach within the framework of many-body perturbation theory, we perform a case study on the dielectric screening effects of hexagonal boron nitride (hBN) on various transition-metal dichalcogenides (TMDs). We consider three systems: monolayer MoS$_2$, bilayer MoS$_2$, and mixed WS$_2$/MoS$_2$ bilayer adsorbed on hBN, and examine three aspects of the substrate dielectric screening: (i) thickness dependence and the monolayer-to-bulk transition, where we consider the effects of one-, two-, three-, and four-layer hBN; (ii) local-field effect, where we numerically assess a common approximation of neglecting the in-plane local-field components of the substrate polarizability; and (iii) spatial dependence, where we consider mixed WS$_2$/MoS$_2$ bilayer adsorbed on hBN with either side facing the substrate. Our results provide quantitative insight into how the substrate screening effects can be leveraged for band structure engineering.
\end{abstract}

 \maketitle

\section{Introduction}

Transition-metal dichalcogenides (TMDs) such as MoS$_2$ and WS$_2$ exhibit unique electronic and optical properties due to the quantum confinement as a result of their low dimensionality \cite{Splendiani2010,Chhowalla2013,Jariwala2014,Qiu2016}. Experimentally, such two-dimensional (2D) materials are often fabricated on substrates via chemical vapor deposition \cite{Wang2015,Ji2015}. TMD-based devices, such as field-effect transistors \cite{Sebastian2021,Liu2013,Podzorov2004}, are often operated with the support of a substrate, too. The coupling between the TMD adsorbates and the supporting substrates is usually van der Waals, indicating little orbital hybridization. However, due to the long-range nature of the Coulomb interaction, it has been extensively shown that the substrates act as dielectric media and effectively screen the Coulomb interaction within the adsorbate \cite{Ryou2016,Hill2016,Qiu2017,Raja2017}, resulting in renormalizations in the fundamental (transport) band gaps and optical excitation energies of the adsorbate \cite{Zheng2016,Cho2018,Zibouche2021}. 

For TMDs, commonly used substrates include Au \cite{Lattyak2022,Panasci2021,Soerensen2014}, graphene/graphite \cite{Liu2016,Huang2015,Chiu2015,Zhang2014}, SiO$_2$/Si \cite{Li2014,Lu2014,Chow2015}, and hexagonal boron nitride (hBN) \cite{Wang2015,Han2019,Zhang2019}. Since large-gap insulators are generally believed to have low dielectric constants \cite{Penn1962}, one might naively assume that SiO$_2$ and hBN (both having band gaps over 6 eV) do not significantly alter the electronic and optical properties of the materials adsorbed on them. However, this has been shown to be not the case. For instance, Ref. \cite{Klein2019} showed that both SiO$_2$ and hBN substrates could considerably modulate the local electronic structure and optical properties of a monolayer MoS$_2$.

Due to the wide use of substrates, it is somewhat challenging to extract the properties of a ``freestanding'' TMD monolayer in experiments and quantify the effects of the substrates. For this purpose, first-principles calculations play an indispensable role. It is now common knowledge that one needs many-body perturbation theory to accurately describe the quasiparticle and excitonic properties of 2D materials, due to deficiencies in typical functionals \cite{Kuemmel2008,Perdew1983,Yang2012} within the framework of density functional theory (DFT). First-principles $GW$-BSE \cite{Hedin1965,Strinati1982,Hybertsen1986,Rohlfing2000} ($G$: Green's function; $W$: screened Coulomb interaction; BSE: Bethe-Salpeter equation) has shown to be essential in quantitative descriptions of the quasiparticle and optical properties of pristine TMDs \cite{Shi2013,Qiu2013,Ramasubramaniam2012,Cheiwchanchamnangij2012}, where local and semi-local functionals in the DFT framework significantly underestimate the band gaps of 2D materials. As far as their interfaces with substrates are concerned, due to the presence of supercells, various approximations have been made to keep the computational cost at a manageable level. Among others, the substrate screening approximation \cite{Ugeda2014} captures the dielectric screening from the substrate \cite{Neaton2006,Thygesen2009} and has proven to be successful for weakly coupled (without significant orbital hybridization) interfaces \cite{Liu2019,Xuan2019,Frimpong2021}. Within the framework of substrate screening, additional approximations are often made to further reduce the computational cost, including the neglect of the local-field components in the non-interacting polarizability or response function ($\chi^0$) of the substrate along directions parallel to the surface \cite{Ugeda2014} and the use of model dielectric functions for the substrate \cite{Zibouche2021,Drueppel2017}. However, the numerical consequences of such additional approximations have not been fully examined, which we would like to address in this work.

In this work, we quantitatively study the effect of hBN substrate on three adsorbates: monolayer MoS$_2$, bilayer MoS$_2$, and the WS$_2$/MoS$_2$ bilayer heterostructure. To focus on the dielectric effect of the substrate on modulating the electronic structure of the adsorbates, we employ the dielectric embedding $GW$ approach \cite{Liu2020}, which yields quantitative agreement with direct $GW$ calculations of the interface, as we explicitly show below. Using monolayer MoS$_2$ adsorbed on few-layer hBN as an example, we first illustrate how the dielectric effect of the substrate grows as the thickness of the hBN, and conclude that it converges at four layers. Then we numerically assess a commonly used approximation, namely the neglect of the local-field components in $\chi^0_{\rm hBN}$ along directions parallel to the surface. After that, we study bilayer MoS$_2$ adsorbed on four-layer hBN and show how the substrate lifts the degeneracy of the bilayer MoS$_2$ bands. Lastly, we study the WS$_2$/MoS$_2$ bilayer adsorbed on four-layer hBN and furthermore compare two scenarios where either the MoS$_2$ or the WS$_2$ side faces the hBN substrate upon adsorption. We demonstrate that the substrate screening is spatially dependent and can be leveraged to selectively modulate certain bands of the WS$_2$/MoS$_2$ bilayer.

The paper is structured as follows. In Sec. \ref{sec:ii}, we describe the computational details and report parameters adopted in our calculations. In Sec. \ref{sec:iii}, the main results are presented in four subsections: (a) the thickness-dependent dielectric screening of hBN, in the context of monolayer MoS$_2$ adsorbed on few-layer hBN; (b) the numerical assessment of the in-plane local-field effects in the $\chi^0_{\rm hBN}$; (c) the case of bilayer MoS$_2$ adsorbed on four-layer hBN and the related degeneracy lifting; and (d) WS$_2$/MoS$_2$ bilayer adsorbed on four-layer hBN, where the spatial dependence of the substrate dielectric effect is discussed. We make concluding remarks in Sec. \ref{sec:iv}.

\section{Computational Details}
\label{sec:ii}
\begin{figure}[htp]
\centering
\includegraphics[width=3.3in]{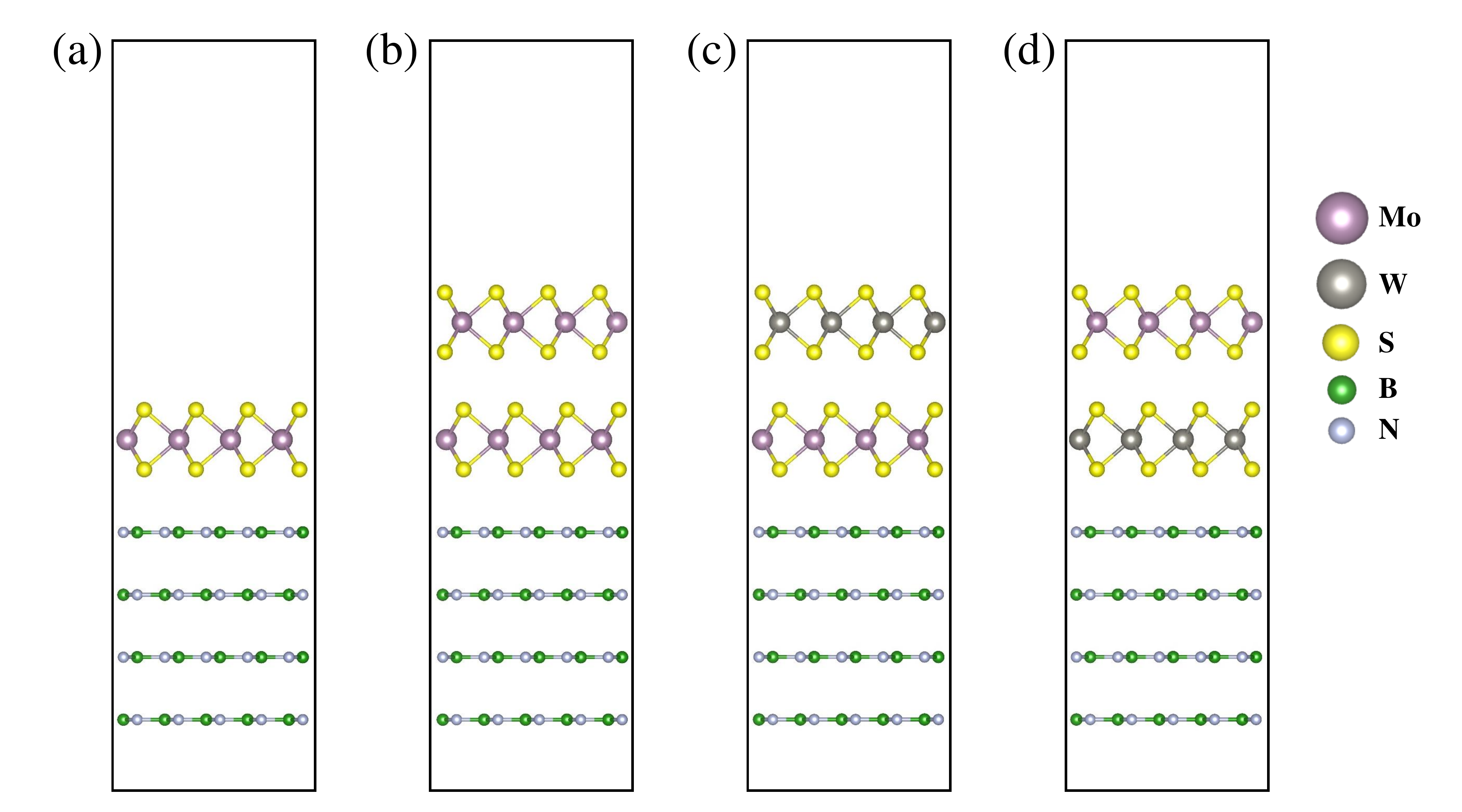}
\caption{The side view of the systems studied in this work. (a) Monolayer MoS$_2$ adsorbed on four-layer hBN (interfaces with one-, two-, and three-layer hBN are not shown); (b) Bilayer MoS$_2$ adsorbed on four-layer hBN; (c) WS$_2$/MoS$_2$ bilayer adsorbed on hBN, with MoS$_2$ being closer to the substrate; (c) WS$_2$/MoS$_2$ bilayer adsorbed on hBN, with WS$_2$ being closer to the substrate. The black boxes represent periodic boundary conditions. The amount of vacuum is not shown in scale. This figure is rendered using VESTA \cite{Momma2008}.}
\label{fig:stru}
\end{figure}

We adopt the lattice parameters and atomic coordinates of monolayer MoS$_2$ unit cell from our previous work \cite{Adeniran2021}, where the in-plane lattice parameter is 3.15 \AA, relaxed using the vdw-DF-cx functional \cite{Berland2014}. To explicitly assess the in-plane local-field effect in $\chi^0_{\rm hBN}$, we need to construct a commensurate supercell for the interface formed between MoS$_2$ and hBN. To minimize the strain on hBN, we construct the supercell containing $4{\times}4$ MoS$_2$ unit cells and $5{\times}5$ hBN unit cells. The resulting hBN in-plane lattice parameter is 2.52 \AA, corresponding to a 0.8\% expansive strain compared to the experimental lattice parameter of 2.50 \AA~\cite{Pease1952}. The internal coordinates of hBN are relaxed using the vdw-DF-cx functional. We have checked that the band structure of hBN is not qualitatively affected by this strain. The length of the unit cell along the out-of-plane direction is 45 \AA, allowing for sufficient vacuum. Multi-layer hBN systems are modeled by placing additional hBN layer(s) 3.3 \AA~apart from one another, consistent with Ref. \cite{Wickramaratne2018} and experiment \cite{Pease1952}. For bilayer MoS$_2$ and the hetero-bilayer WS$_2$/MoS$_2$, we adopt the AA' stacking and fully relax the atomic coordinates using the vdw-DF-cx functional until all residual forces are below 0.05 eV/{\AA}. We optimize the adsorption height of each adsorbate on the hBN substrate using the vdw-DF-cx functional without further relaxation of the internal atomic coordinates of the adsorbate or substrate, resulting in about 3.3 \AA~adsorption height for each interface. All DFT calculations employ the optimized norm-conserving Vanderbilt (ONCV) pseudopotentials \cite{Schlipf2015,Hamann2013} and the \texttt{Quantum ESPRESSO} package \cite{Giannozzi2017}. Fig. \ref{fig:stru} shows all systems studied in this work.

For the $GW$ calculations, we employ the dielectric embedding $GW$ scheme \cite{Liu2020} to focus on the dielectric effects of the hBN substrate on the TMD adsorbates. Using MoS$_2$ adsorbed on mono-layer hBN as an example, we have checked explicitly that the results from the dielectric embedding $GW$ agree well with those obtained from a direct $GW$ calculation of the MoS$_2$:hBN interface system. In the dielectric embedding $GW$ approach, the dielectric function is determined from a combined $\chi^0$, i.e., $\chi^0(\mathbf{q})=\chi^0_{\rm TMD}(\mathbf{q})+\chi^0_{\rm hBN}(\mathbf{q})$. Physically, this approximation in $\chi^0$ neglects the orbital hybridization or mixing between TMD and hBN, which is a good approximation given the fact that the interaction between the two materials is mainly van der Waals. Both the $\chi^0_{\rm TMD}$ and the $\chi^0_{\rm hBN}$ are first calculated in their corresponding unit cells and then folded in the reciprocal space into the interface supercells. For both subsystems, we use a 5 Ry dielectric cutoff and 200 bands in the summation to compute $\chi^0$, and 30 bands on a shifted $\mathbf{k}$-grid for the treatment of the $\mathbf{q} \to 0$ limit. A $\mathbf{q}$-mesh of $20{\times}20{\times}1$ ($25{\times}25{\times} 1$) is used for the TMD (hBN) unit cell. The resulting combined $\chi^0$ for the interface supercell is then expressed on a $\mathbf{q}$-mesh of $5\times5\times1$.

In the dielectric embedding $GW$ approach, the self-energy calculations are performed for the supercells of TMD adsorbates in the original interface simulation cells without explicitly including the substrate atoms. Physically, this approach neglects the overlap of TMD and hBN orbitals in the expression for the self-energy \cite{Liu2020}, consistent with the approximation in $\chi^0$ discussed above. The calculations use a $\mathbf{k}$-mesh of $5{\times}5{\times}1$, a 5 Ry dielectric cutoff, and 4800 bands in the summation for the Green's function. All $GW$ calculations are at the $G_0W_0$ level as implemented in the \texttt{BerkeleyGW} package \cite{Deslippe2012}, using the Perdew-Burke-Ernzerhof (PBE) functional \cite{Perdew1996} as the mean-field starting point. The frequency dependence of the dielectric function is treated using the Hybertsen-Louie generalized plasmon-pole model \cite{Hybertsen1986}, with the plasma frequency in the embedding $GW$ calculation of TMD set to be that from the corresponding TMD:hBN interface \cite{Liu2020}. The $\mathbf{q}\to 0$ limit is treated using the semiconductor screening, and the spurious long-range interactions along the out-of-plane direction are removed by employing the slab Coulomb truncation \cite{IsmailBeigi2006}. The static remainder \cite{Deslippe2013} is used to improve the convergence of the self-energy. All $GW$ band structures are obtained via interpolation of the quasiparticle corrections from the explicitly computed $5\times5\times1$ $\mathbf{k}$-mesh to uniformly sampled $\mathbf{k}$ points along the high-symmetry lines in the Brillouin zone, as implemented in the \texttt{BerkeleyGW} package \cite{Deslippe2012}.

\section{Results and Discussion}
\label{sec:iii}
\subsection{Monolayer MoS$_2$ on few-layer hBN}
We first study how the dielectric screening effect of the hBN substrate varies as the number of hBN layers increases. It will provide insight into the difference between two common scenarios explored in experiments and modelling, i.e., a ``one-layer hBN substrate'' \cite{Trolle2017,Bruelke2020} versus a ``few-layer hBN substrate'' \cite{Klein2019,Wang2015}. Furthermore, it will indicate the minimum number of hBN layers needed in computational modeling to mimic the dielectric effect of a semi-infinite hBN substrate on an adsorbate. 

\begin{figure}[htp]
\centering
\includegraphics[width=3.3in]{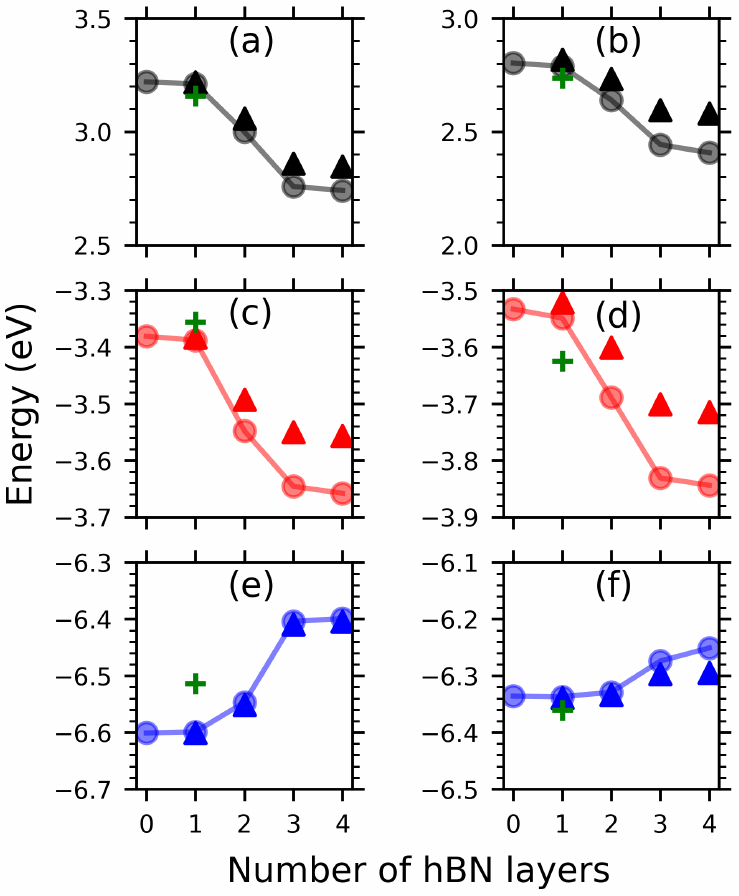}
\caption{Comparison of the band gap (black), the CBM energy (red), and the VBM energy (blue) at two different $\mathbf{k}$ points for monolayer MoS$_2$ adsorbed on few-layer hBN. Zero-layer hBN indicates a freestanding MoS$_2$ monolayer. All energies are measured with respect to the vacuum. (a), (c), and (e) are for the $\Gamma$ point, while (b), (d), and (f) are for the $\mathbf{k}$ point with fractional coordinates (0.4,0.4,0.0), which is an explicitly computed $\mathbf{k}$ point closest to K, where the direct band gap of monolayer MoS$_2$ lies. Circles (triangles) represent results from the dielectric embedding $GW$ calculations of MoS$_2$ with (without) the in-plane local-field effect of $\chi^0_{\rm hBN}$. Green cross markers represent results from a direct $GW$ calculation of an interface formed between monolayer MoS$_2$ and monolayer hBN.}
\label{fig:ml}
\end{figure}

\begin{figure*}[htp]
\centering
\includegraphics[width=6.5in]{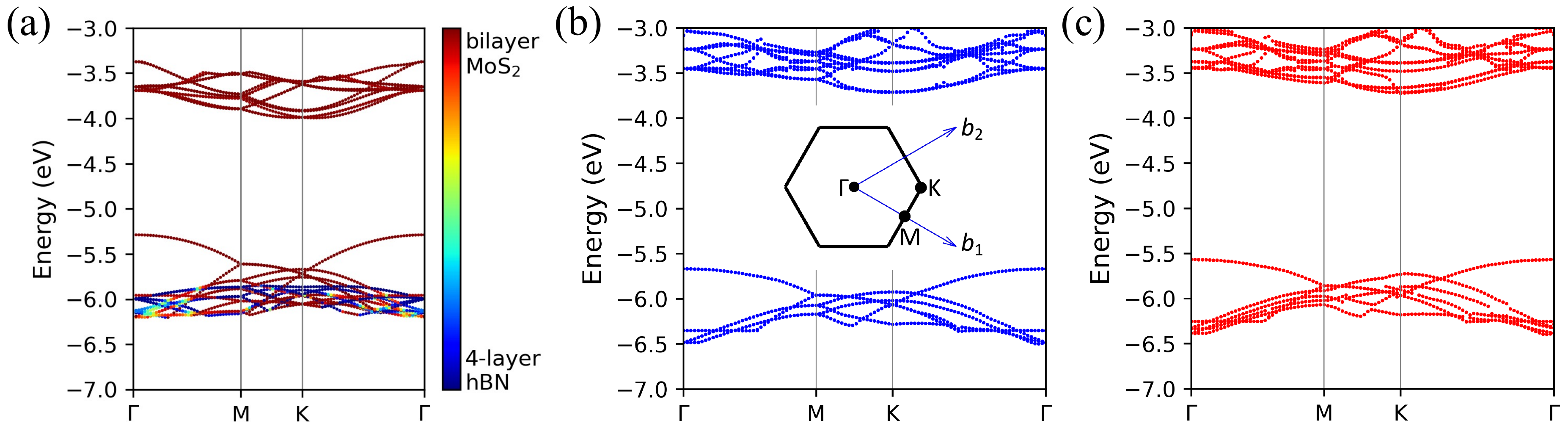}
\caption{(a) Color-coded PBE band structure of the bilayer MoS$_2$ adsorbed on four-layer hBN. (b) $GW$ band structure of freestanding bilayer MoS$_2$. The inset shows the Brillouin zone and the high-symmetry points, as well as the in-plane reciprocal-space lattice vectors $\vec{b}_1$ and $\vec{b}_2$. (c) Band structure from a dielectric embedding $GW$ calculation of bilayer MoS$_2$ adsorbed on four-layer hBN. All calculations are carried out in a 4$\times$4 supercell of bilayer MoS$_2$ (commensurate with 5$\times$5 hBN supercell) and all energies are measured with respect to the vacuum level. The $GW$ band structures in (b) and (c) are obtained by interpolating explicitly computed 25 valence bands (24 conduction bands) on the 5$\times5$ $\mathbf{k}$-mesh to 6 valence bands (15 conduction bands) on 100 uniformly sampled $\mathbf{k}$-points along the high symmetry lines. The seemingly ``discontinuous'' bands seen in (b) and (c) are likely due to the perturbative nature of $G_0W_0$ and the specific energy window used in the interpolation.}
\label{fig:bl}
\end{figure*}

Fig. \ref{fig:ml} illustrates how the band gap [(a)(b)], conduction band minimum (CBM) energies [(c)(d)], and valence band maximum (VBM) energies [(e)(f)] vary as a function of the number of hBN layers, at two different $\mathbf{k}$ points in the Brillouin zone: $\Gamma$ [(a)(c)(e)] and $(0.4,0.4,0.0)$ [(b)(d)(f)], where the latter is expressed in fractional coordinates (measured with respect to the lattice basis vectors) in the reciprocal space. The Brillouin zone and in-plane reciprocal-space lattice vectors are shown in the inset in Fig. \ref{fig:bl}(b), which applies to all systems in this work. Monolayer MoS$_2$ is a direct gap material \cite{Mak2010} with the band gap at the K point, and the direct band gap remains at the K point when MoS$_2$ is deposited on the hBN substrate. We report information at the $\mathbf{k}$ point of $(0.4,0.4,0.0)$ simply because it is closest to K among all the directly calculated $\mathbf{k}$ points using $GW$ (such that a numerical interpolation is not needed for Fig. \ref{fig:ml}). In each panel, the circles represent results from dielectric embedding $GW$ calculations of MoS$_2$ (including in-plane local-field effects of $\chi^0_{\rm hBN}$, see Sec. \ref{sec:lf} below for further details), with the vacuum level set to zero. Green cross markers represent results from a direct $GW$ calculation of the monolayer MoS$_2$:monolayer hBN interface, as a benchmark for our dielectric embedding $GW$ calculations. One can see that the dielectric embedding $GW$ agrees with direct $GW$ calculations within 0.1 eV for individual energy levels and within 0.05 eV for the gap.

\begin{table*}
\centering
\caption{Comparison of different experimental and theoretical approaches for the influence of the hBN substrate on the monolayer MoS$_2$ band gap (in eV). The MoS$_2$ gap listed in this table for this work is the value at the $\mathbf{k}$-point of (0.4,0.4,0) [Fig. \ref{fig:ml}(b)], which should be numerically similar to the gap at the K point, given the small band curvature at K.}
\begin{tabular}{c|c|c|c}
\hline\hline
Method & Description of the hBN Dielectric Effect & MoS$_2$ Gap & Reference \\
\hline
Sternheimer $GW$, Godby-Needs plasmon-pole &  model $W$, $\epsilon_{\rm eff}=(1+\epsilon_{\rm S})/2$, $\epsilon_{\rm S}=2.6$ & 2.58 & \cite{Zibouche2021} \\
Sternheimer $GW$, full frequency & model $W$, $\epsilon_{\rm eff}=(1+\epsilon_{\rm S})/2$, $\epsilon_{\rm S}=2.6$ & 2.50 & \cite{Zibouche2021} \\

eigenvalue-self-consistent $GW_0$, full frequency & similar to Eq. \eqref{eq:sswo}, extrapolated from a 14-layer & 2.36 & \cite{Utama2019} \\
& hBN to semi-infinite thickness & & \\
LDA+$GdW$, $dW=W-W_{\rm metal}$ \cite{Rohlfing2010} &  models for $W$ and $W_{\rm metal}$, $\epsilon_{\infty}^{\parallel}/\epsilon_{\infty}^{\perp}=4.95/4.10$ & 2.73 & \cite{Drueppel2017} \\
LDA+$G_0W_0$ & monolayer hBN, similar to Eq. \eqref{eq:sswo} & 2.66 & \cite{Naik2018} \\
scanning tunneling spectroscopy & experiment & 2.35 & \cite{Klein2019} \\
Dielectric embedding $G_0W_0$ & Comparing Eq. \eqref{eq:ss} and Eq. \eqref{eq:sswo}, 4-layer hBN & 2.41 & This Work \\
\hline
\end{tabular}
\label{tab:compare}
\end{table*}

Fig. \ref{fig:ml} shows interesting trends as the thickness of hBN substrate grows. We first notice that one-layer hBN barely provides any dielectric screening to the monolayer MoS$_2$ adsorbate, as the VBM, CBM, and gap values do not change considerably compared to the freestanding monolayer MoS$_2$. Physically, this can be attributed to the fact that one-layer hBN is atomically thin, such that the adsorbate can not induce a strong density response perpendicular to the surface ``within'' the substrate that is responsible for the dielectric screening. This might explain why single-layer hBN can be used as a ``spacer'' that effectively decouples an adsorbate from a substrate (see, e.g., Ref. \cite{Rizzo2020} for an example in experiment). The dielectric screening becomes stronger as the number of hBN layers increases, and eventually saturates when the number of hBN layers reaches four. Therefore, we conclude that for the purpose of modeling a semi-infinite hBN substrate, four layers are enough to converge the results within 0.1 eV, as far as the band gap of the adsorbate is concerned. This conclusion is in line with Ref. \cite{Wickramaratne2018}, where the monolayer-to-bulk transition of hBN band structure is analyzed in detail.

It is worth mentioning that the dielectric embedding $GW$ approach used here is particularly efficient for the purpose of studying the thickness dependence of the substrate dielectric screening. This is because we can bypass the calculation of the wavefunctions and dielectric functions of a series of large interfaces. Instead, we only need to compute the $\chi^0_{{\rm MoS}_2}$ once and combine it with the $\chi^0_{\rm hBN}$ calculated for different layers of the substrate, with both components explicitly calculated in their unit cells rather than supercells. Such a procedure can be easily generalized to analyze the effects of other substrates. 

We compare our results to existing literature. Table \ref{tab:compare} compares prior experimental and theoretical studies of the dielectric screening effect of hBN on monolayer MoS$_2$. In particular, we list how each computational method treats the dielectric effect of the hBN substrate, either from first-principles \cite{Utama2019,Naik2018} or from model dielectric functions \cite{Zibouche2021,Drueppel2017}. Our full first-principles $GW$ results are consistent with other similar studies and experiments, and provide another benchmark for developing model dielectric functions.

Lastly, we note in passing that we focus on the gap of the monolayer MoS$_2$, rather than the band alignment between MoS$_2$ and hBN. DFT calculations indicate that the MoS$_2$:hBN interfaces are type-I, in the sense that the band edges of the MoS$_2$ are sandwiched between those of the hBN, and the VBM/CBM of the interfaces are entirely localized on the MoS$_2$. While the VBM of the hBN substrate is within 1 eV (DFT result) below that of the MoS$_2$, the CBM of the hBN substrate is many eVs above that of MoS$_2$. These energy level offsets are believed to be even larger in $GW$ \cite{Zibouche2021,Drueppel2017,Utama2019}, such that the charge transfer between the two components is negligible. 

\subsection{In-plane local-field effect of substrate $\chi^0$}
\label{sec:lf}

Pioneered by Ref. \cite{Ugeda2014}, a common approximation in the study of substrate dielectric effect is to neglect the in-plane (parallel to the surface) local-field components of the substrate $\chi^0$, i.e., 
\begin{equation}
\bar{\chi}^0_{\rm hBN}(\mathbf{G},\mathbf{G'};\mathbf{q}) = \chi^0_{\rm hBN}(\mathbf{G},\mathbf{G'};\mathbf{q})\delta_{G_xG'_x}\delta_{G_yG'_y},
\label{eq:lfe}
\end{equation}
where $\chi^0_{\rm hBN}$ is the directly calculated $\chi^0$ using the random-phase approximation, and $\bar{\chi}^0$ neglects the in-plane local-field components of $\chi^0_{\rm hBN}$. Subsequently, in the substrate screening approximation:
\begin{equation}
\chi^0_{\rm tot}\approx \chi^0_{\rm{MoS}_2}+\chi^0_{\rm hBN}, \hspace{0.1in} \mbox{with local-field effect;} 
\label{eq:ss}
\end{equation}
\begin{equation}
\bar{\chi}^0_{\rm tot}\approx \chi^0_{\rm{MoS}_2}+\bar{\chi}^0_{\rm hBN},  \hspace{0.1in} \mbox{without local-field effect.}
\label{eq:sswo}
\end{equation}
Note that the local-field components of the adsorbate, i.e., those in $\chi^0_{\rm{MoS}_2}$, are always included. The neglect of the substrate local-field effect enables calculations using incommensurate supercells between the adsorbate and the substrate \cite{Ugeda2014} and greatly reduces the computational cost. However, to the best of our knowledge, its accuracy has not been explicitly examined. In this work, we assess the numerical consequence of Eq. \eqref{eq:lfe} for the first time, in the context of monolayer MoS$_2$ adsorbed on few-layer hBN.

The results are shown in Fig. \ref{fig:ml}. Circles (triangles) represent results from Eq. \eqref{eq:ss} [Eq. \eqref{eq:sswo}], i.e., with (without) the in-plane local-field components of $\chi^0_{\rm hBN}$. We can see that the local-field effect in the dielectric screening increases as the number of hBN layers. For the four-layer hBN substrate, the neglect of the local-field components in $\chi^0_{\rm hBN}$ causes an error of 0.2 eV in the MoS$_2$ gap [Fig. \ref{fig:ml}(b)], which can be decomposed to a 0.15 eV error in the CBM [Fig. \ref{fig:ml}(d)] and another 0.05 eV error in the VBM [Fig. \ref{fig:ml}(f)]. Interestingly, we find that the local-field effect is larger at the $\mathbf{k}$ point of (0.4,0.4,0.0) than at $\Gamma$, and is larger for CBM than for VBM. This is likely due to the different shapes of the wavefunctions of the pertinent energy levels. 

As a concluding remark, the neglect of local-field effects leads to a moderate error of 0.2 eV in the MoS$_2$ gap in the current study. However, the magnitude of the error may vary for other substrates or adsorbates. Therefore, we believe more systematic studies are needed to fully assess the numerical consequences of the local-field effects in the substrate dielectric screening.


\subsection{Bilayer MoS$_2$ on four-layer hBN}
Having established that the dielectric screening effect of the hBN substrate converges at four layers, we only consider TMDs adsorbed on four-layer hBN in the subsequent calculations.

Fig. \ref{fig:bl}(b)(c) compare the $GW$ band structure of the freestanding bilayer MoS$_2$ and when it is adsorbed on four-layer hBN, where the latter is computed using the dielectric embedding $GW$ approach. For the freestanding bilayer, we obtain a value of 2.0 eV for the quasiparticle gap, which is in good agreement with other $GW$ calculations (1.9 eV from Ref. \cite{Cheiwchanchamnangij2012}). For the MoS$_2$ adsorbed on four-layer hBN, the gap is reduced to 1.8 eV. Here, the gap renormalization of 0.2 eV is smaller than that of the monolayer MoS$_2$ [0.4 eV, see Fig. \ref{fig:ml}(b)]. This is because, in bilayer MoS$_2$, the orbitals are delocalized over both layers, with electron distribution moving further away from the substrate than in the case of monolayer MoS$_2$, resulting in weaker substrate dielectric screening.

Furthermore, we note that the dielectric screening of the hBN substrate enhances the degeneracy lifting of some of the bilayer MoS$_2$ bands, in addition to that captured at the PBE level. Fig. \ref{fig:bl}(a) shows the color-coded PBE band structure for the interface formed between bilayer MoS$_2$ and four-layer hBN. Comparing it with the $GW$ band structure of the freestanding bilayer MoS$_2$ in Fig. \ref{fig:bl}(b), one finds that some degeneracies of the bilayer MoS$_2$ bands are lifted by the crystal field effect of the hBN, such as the CBM at the $\Gamma$ point. However, the degeneracy lifting captured by PBE is not so pronounced for other bands, such as the CBM at the K point, a few inner valence bands at the $\Gamma$ point, [around -4.5 eV in Fig. \ref{fig:bl}(a) and -6.5 eV in Fig. \ref{fig:bl}(b)], and many other near-degenerate cases in the Brillouin zone. In fact, at the PBE level of theory, for those near-degenerate bands, the degeneracy lifting due to the substrate is only on the order of 10$^{-6}$ eV. Notably, the substrate dielectric screening greatly enhances the degeneracy lifting, causing a band splitting on the order of 10 meV, which can be seen in the embedding $GW$ results in Fig. \ref{fig:bl}(c). 

Comparing the bilayer MoS$_2$ band structure between its freestanding form and after it is adsorbed on hBN, one finds that the former features many degeneracies, due to the symmetry of the two MoS$_2$ layers. The symmetry is broken and the degeneracies are lifted when the bilayer MoS$_2$ is adsorbed on substrates such as the four-layer hBN considered here. It is still an open question whether or not the approximate electronic structure methods in use (either PBE or $G_0W_0$) could preserve certain symmetries that are protected and break others that should be broken in the interface. We find that the specific density functional used here, namely the PBE, is able to lift some degeneracies when MoS$_2$ is adsorbed on hBN, i.e., capturing the orbital mixing between MoS$_2$ and hBN that is part of the crystal field effects. However, PBE seems to have difficulty in completely lifting the degeneracies in other bands and $\mathbf{k}$ points (only causing a band splitting of about $10^{-6}$ eV, as we discussed above). For those near-degenerate bands in PBE, the embedding $GW$ approach is able to further break the symmetry and enhance the degeneracy lifting, causing a band splitting on the order of 10 meV. The latter is physically a result of the dielectric screening of the substrate, because one layer is in closer contact with the substrate and feels stronger dielectric screening than the other.  

\begin{figure*}[htb]
\centering
\includegraphics[width=6.5in]{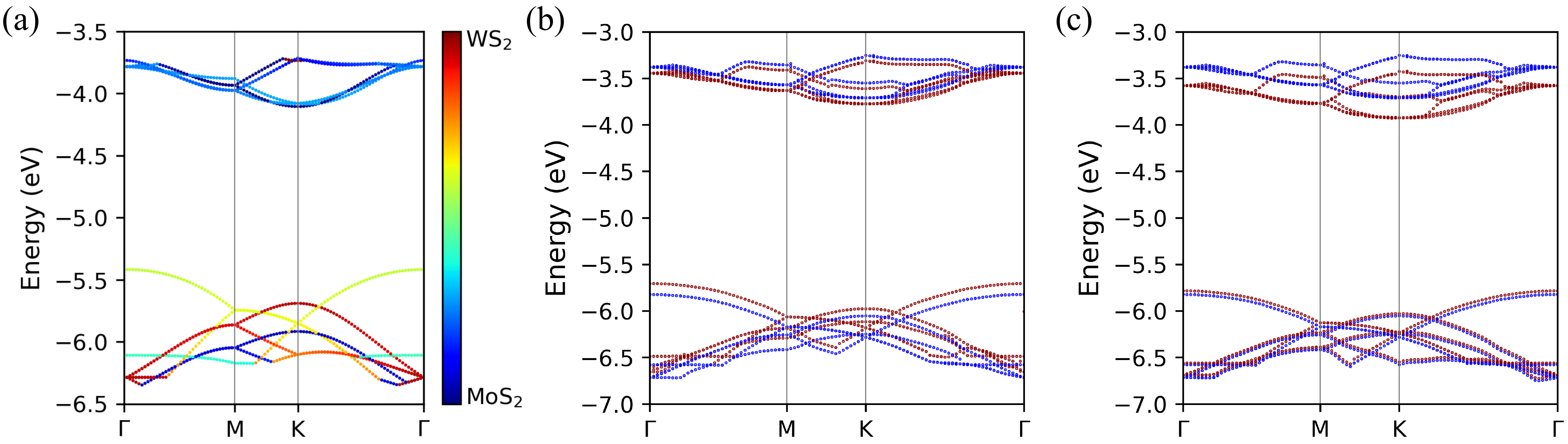}
\caption{(a) Color-coded PBE band structure of the freestanding WS$_2$/MoS$_2$ heterobilayer. In both (b) and (c), blue dots denote the $GW$ band structure of the freestanding WS$_2$/MoS$_2$ heterobilayer and red dots denote the band structure from dielectric embedding $GW$ calculations of the WS$_2$/MoS$_2$ heterobilayer. In (b) [(c)], the WS$_2$ (MoS$_2$) side is closer to the four-layer hBN substrate, as shown in Fig. \ref{fig:stru}(d) [Fig. \ref{fig:stru}(c)]. All calculations are carried out in a 4$\times$4 supercell of WS$_2$/MoS$_2$ heterobilayer (commensurate with 5$\times$5 hBN supercell) and all energies are measured with respect to the vacuum level.}
\label{fig:mb}
\end{figure*}

\subsection{Mixed WS$_2$/MoS$_2$ bilayer on four-layer hBN}

In this subsection, we examine the dielectric effect of the hBN substrate on the electronic structure of mixed WS$_2$/MoS$_2$ bilayer, and show that different placements of the hBN lead to different modulations of the WS$_2$/MoS$_2$ bilayer band structure.

Figs. \ref{fig:mb}(b) and \ref{fig:mb}(c) compare the $GW$ band structure of the WS$_2$/MoS$_2$ bilayer in its freestanding phase and when it is adsorbed on the four-layer hBN substrate. In both panels, blue dots denote the $GW$ band structure of the freestanding WS$_2$/MoS$_2$ bilayer (identical in both panels), where we obtain a 2.11 eV for the gap between $\Gamma$ and K (compared to 1.96 eV in Ref. \cite{Debbichi2014}), and a 2.50 eV for the direct band gap at K (compared to 2.42 eV in Ref. \cite{Gao2017}). In Fig. \ref{fig:mb}(b) [Fig. \ref{fig:mb}(c)], the red dots denote the band structure computed from dielectric embedding $GW$, taking into account the dielectric screening effects from the four-layer hBN that is placed close to the WS$_2$ (MoS$_2$) side, as shown in Fig. \ref{fig:stru}(d) [Fig. \ref{fig:stru}(c)]. We first note that the qualitative features of the WS$_2$/MoS$_2$ bands are preserved upon adsorption on hBN. The band gap in Fig. \ref{fig:mb}(b) is 1.93 eV, compared to 1.86 eV in Fig. \ref{fig:mb}(c). We note that this difference in the band gap is consistent with the trend in the dielectric function of the embedded WS$_2$/MoS$_2$ system: the $\epsilon_{\mathbf{G}=0,\mathbf{G}'=0}(\mathbf{q})$ is slightly smaller for the case in Fig. \ref{fig:mb}(b) for every $\mathbf{q}$ point in the Brillouin zone ($\mathbf{G}$ and $\mathbf{G}$' are reciprocal-space lattice vectors). Physically, this can be attributed to the slightly smaller in-plane dielectric constant in WS$_2$ compared to MoS$_2$ \cite{Laturia2018}.

Crucially, the valence band energies are renormalized by the substrate to a greater extent than the conduction band energies (compare red dots with blue dots) in Fig. \ref{fig:mb}(b), and the scenario is the opposite in Fig. \ref{fig:mb}(c). To understand this phenomenon, we identify the nature of the valence and conduction bands of the freestanding WS$_2$/MoS$_2$ bilayer. In Fig. \ref{fig:mb}(a), we color-code the PBE band structure of the freestanding WS$_2$/MoS$_2$ bilayer, where the top valence band is localized on WS$_2$ and the bottom conduction band is localized on MoS$_2$. These orbital assignments are consistent with literature \cite{Bernardi2013}. When the mixed WS$_2$/MoS$_2$ bilayer is adsorbed on the hBN substrate with WS$_2$ (MoS$_2$) side facing the substrate, the dielectric screening of the hBN has a higher impact on the valence (conduction) bands of the mixed WS$_2$/MoS$_2$ bilayer, due to its spatial closeness to WS$_2$ (MoS$_2$). This spatial-dependent dielectric screening is reflected in the band structure calculated from the dielectric embedding $GW$, i.e., the red dots in Fig. \ref{fig:mb}(b) and Fig. \ref{fig:mb}(c). In conclusion, adsorbate orbitals localized closer to the substrate experience stronger screening from the substrate, leading to larger energy renormalization than those localized farther away from the substrate.

We note that Ref. \cite{Hill2016} reported different band onset energies for two vertical stacking configurations: MoS$_2$/WS$_2$/quartz and WS$_2$/MoS$_2$/quartz, based on scanning tunneling spectroscopy results. Although our work focuses on a different substrate, our results qualitatively explain the phenomenon observed in Ref. \cite{Hill2016}, in terms of the spatial dependence of the substrate dielectric screening. We further comment that the dielectric embedding $GW$ approach is capable of capturing this effect efficiently. The placement of the substrate provides an effective way to selectively tune the band gaps of complex heterostructures.

\section{Conclusions}
\label{sec:iv}
In this work, we have carried out dielectric embedding $GW$ calculations to scrutinize the dielectric screening effects of hBN substrate on the electronic structure of TMD adsorbates. We considered three sets of systems: (i) monolayer MoS$_2$ adsorbed on one-, two-, three-, and four-layer hBN; (ii) bilayer MoS$_2$ adsorbed on four-layer hBN; and (iii) mixed WS$_2$/MoS$_2$ bilayer adsorbed on four-layer hBN with either WS$_2$ or MoS$_2$ facing the substrate. Our main findings are the following: (i) monolayer hBN provides nearly negligible screening to the adsorbate, and the dielectric screening effect of the hBN substrate converges at four layers; (ii) a common approximation used, i.e., neglect of the in-plane local-field components of the hBN polarizability ($\chi^0_{\rm hBN}$) can introduce an error of about 0.2 eV in the gap of the TMD adsorbate; (iii) the dielectric effect of the substrate enhances the symmetry breaking of the bilayer MoS$_2$, in addition to the crystal field effect; and (iv) different placements of the substrate can be used to selectively modulate the valence or conduction bands of mixed WS$_2$/MoS$_2$ bilayer. Additionally, we have found that the gap of the mixed WS$_2$/MoS$_2$ bilayer is slightly smaller when the MoS$_2$ side is placed closer to the substrate. Overall, the qualitative trends derived from our work provide insights into substrate-induced properties and substrate-based band structure engineering, and we hope our work provides a paradigm for future studies of interfaces with physisorption.

\section{Acknowledgements}
We thank Li Yang for fruitful discussions. Z.-F.L. acknowledges an NSF CAREER Award, DMR-2044552. O.A. acknowledges a summer dissertation fellowship and a Rumble fellowship and the A. Paul and Carole C. Schaap endowed distinguished graduate award in chemistry at Wayne State University. DFT calculations used computational resources at the Center for Functional Nanomaterials (CFN), which is a U.S. Department of Energy Office of Science User Facility, at Brookhaven National Laboratory under Contract No. DE-SC0012704. Large-scale $GW$ calculations used resources at the National Energy Research Scientific Computing Center (NERSC), a U.S. Department of Energy Office of Science User Facility located at Lawrence Berkeley National Laboratory, operated under Contract No. DE-AC02-05CH11231 through NERSC award BES-ERCAP0020328, as well as the Extreme Science and Engineering Discovery Environment (XSEDE), which is supported by NSF grant number ACI-1548562 through allocation PHY220043.
\bibliography{hbn.bib}

\begin{thebibliography}{72}%
\makeatletter
\providecommand \@ifxundefined [1]{%
 \@ifx{#1\undefined}
}%
\providecommand \@ifnum [1]{%
 \ifnum #1\expandafter \@firstoftwo
 \else \expandafter \@secondoftwo
 \fi
}%
\providecommand \@ifx [1]{%
 \ifx #1\expandafter \@firstoftwo
 \else \expandafter \@secondoftwo
 \fi
}%
\providecommand \natexlab [1]{#1}%
\providecommand \enquote  [1]{``#1''}%
\providecommand \bibnamefont  [1]{#1}%
\providecommand \bibfnamefont [1]{#1}%
\providecommand \citenamefont [1]{#1}%
\providecommand \href@noop [0]{\@secondoftwo}%
\providecommand \href [0]{\begingroup \@sanitize@url \@href}%
\providecommand \@href[1]{\@@startlink{#1}\@@href}%
\providecommand \@@href[1]{\endgroup#1\@@endlink}%
\providecommand \@sanitize@url [0]{\catcode `\\12\catcode `\$12\catcode
  `\&12\catcode `\#12\catcode `\^12\catcode `\_12\catcode `\%12\relax}%
\providecommand \@@startlink[1]{}%
\providecommand \@@endlink[0]{}%
\providecommand \url  [0]{\begingroup\@sanitize@url \@url }%
\providecommand \@url [1]{\endgroup\@href {#1}{\urlprefix }}%
\providecommand \urlprefix  [0]{URL }%
\providecommand \Eprint [0]{\href }%
\providecommand \doibase [0]{https://doi.org/}%
\providecommand \selectlanguage [0]{\@gobble}%
\providecommand \bibinfo  [0]{\@secondoftwo}%
\providecommand \bibfield  [0]{\@secondoftwo}%
\providecommand \translation [1]{[#1]}%
\providecommand \BibitemOpen [0]{}%
\providecommand \bibitemStop [0]{}%
\providecommand \bibitemNoStop [0]{.\EOS\space}%
\providecommand \EOS [0]{\spacefactor3000\relax}%
\providecommand \BibitemShut  [1]{\csname bibitem#1\endcsname}%
\let\auto@bib@innerbib\@empty
\bibitem [{\citenamefont {Splendiani}\ \emph {et~al.}(2010)\citenamefont
  {Splendiani}, \citenamefont {Sun}, \citenamefont {Zhang}, \citenamefont {Li},
  \citenamefont {Kim}, \citenamefont {Chim}, \citenamefont {Galli},\ and\
  \citenamefont {Wang}}]{Splendiani2010}%
  \BibitemOpen
  \bibfield  {author} {\bibinfo {author} {\bibfnamefont {A.}~\bibnamefont
  {Splendiani}}, \bibinfo {author} {\bibfnamefont {L.}~\bibnamefont {Sun}},
  \bibinfo {author} {\bibfnamefont {Y.}~\bibnamefont {Zhang}}, \bibinfo
  {author} {\bibfnamefont {T.}~\bibnamefont {Li}}, \bibinfo {author}
  {\bibfnamefont {J.}~\bibnamefont {Kim}}, \bibinfo {author} {\bibfnamefont
  {C.-Y.}\ \bibnamefont {Chim}}, \bibinfo {author} {\bibfnamefont
  {G.}~\bibnamefont {Galli}},\ and\ \bibinfo {author} {\bibfnamefont
  {F.}~\bibnamefont {Wang}},\ }\bibfield  {title} {\bibinfo {title} {Emerging
  photoluminescence in monolayer {M}o{S}$_2$},\ }\href
  {https://doi.org/10.1021/nl903868w} {\bibfield  {journal} {\bibinfo
  {journal} {Nano Lett.}\ }\textbf {\bibinfo {volume} {10}},\ \bibinfo {pages}
  {1271} (\bibinfo {year} {2010})}\BibitemShut {NoStop}%
\bibitem [{\citenamefont {Chhowalla}\ \emph {et~al.}(2013)\citenamefont
  {Chhowalla}, \citenamefont {Shin}, \citenamefont {Eda}, \citenamefont {Li},
  \citenamefont {Loh},\ and\ \citenamefont {Zhang}}]{Chhowalla2013}%
  \BibitemOpen
  \bibfield  {author} {\bibinfo {author} {\bibfnamefont {M.}~\bibnamefont
  {Chhowalla}}, \bibinfo {author} {\bibfnamefont {H.~S.}\ \bibnamefont {Shin}},
  \bibinfo {author} {\bibfnamefont {G.}~\bibnamefont {Eda}}, \bibinfo {author}
  {\bibfnamefont {L.-J.}\ \bibnamefont {Li}}, \bibinfo {author} {\bibfnamefont
  {K.~P.}\ \bibnamefont {Loh}},\ and\ \bibinfo {author} {\bibfnamefont
  {H.}~\bibnamefont {Zhang}},\ }\bibfield  {title} {\bibinfo {title} {The
  chemistry of two-dimensional layered transition metal dichalcogenide
  nanosheets},\ }\href {https://doi.org/10.1038/nchem.1589} {\bibfield
  {journal} {\bibinfo  {journal} {Nat. Chem.}\ }\textbf {\bibinfo {volume}
  {5}},\ \bibinfo {pages} {263} (\bibinfo {year} {2013})}\BibitemShut {NoStop}%
\bibitem [{\citenamefont {Jariwala}\ \emph {et~al.}(2014)\citenamefont
  {Jariwala}, \citenamefont {Sangwan}, \citenamefont {Lauhon}, \citenamefont
  {Marks},\ and\ \citenamefont {Hersam}}]{Jariwala2014}%
  \BibitemOpen
  \bibfield  {author} {\bibinfo {author} {\bibfnamefont {D.}~\bibnamefont
  {Jariwala}}, \bibinfo {author} {\bibfnamefont {V.~K.}\ \bibnamefont
  {Sangwan}}, \bibinfo {author} {\bibfnamefont {L.~J.}\ \bibnamefont {Lauhon}},
  \bibinfo {author} {\bibfnamefont {T.~J.}\ \bibnamefont {Marks}},\ and\
  \bibinfo {author} {\bibfnamefont {M.~C.}\ \bibnamefont {Hersam}},\ }\bibfield
   {title} {\bibinfo {title} {Emerging device applications for semiconducting
  two-dimensional transition metal dichalcogenides},\ }\href@noop {} {\bibfield
   {journal} {\bibinfo  {journal} {ACS Nano}\ }\textbf {\bibinfo {volume}
  {8}},\ \bibinfo {pages} {1102} (\bibinfo {year} {2014})}\BibitemShut
  {NoStop}%
\bibitem [{\citenamefont {Qiu}\ \emph {et~al.}(2016)\citenamefont {Qiu},
  \citenamefont {da~Jornada},\ and\ \citenamefont {Louie}}]{Qiu2016}%
  \BibitemOpen
  \bibfield  {author} {\bibinfo {author} {\bibfnamefont {D.~Y.}\ \bibnamefont
  {Qiu}}, \bibinfo {author} {\bibfnamefont {F.~H.}\ \bibnamefont
  {da~Jornada}},\ and\ \bibinfo {author} {\bibfnamefont {S.~G.}\ \bibnamefont
  {Louie}},\ }\bibfield  {title} {\bibinfo {title} {Screening and many-body
  effects in two-dimensional crystals: {M}onolayer {M}o{S}$_2$},\ }\href@noop
  {} {\bibfield  {journal} {\bibinfo  {journal} {Phys. Rev. B}\ }\textbf
  {\bibinfo {volume} {93}},\ \bibinfo {pages} {235435} (\bibinfo {year}
  {2016})}\BibitemShut {NoStop}%
\bibitem [{\citenamefont {Wang}\ \emph {et~al.}(2015)\citenamefont {Wang},
  \citenamefont {Wang},\ and\ \citenamefont {Warner}}]{Wang2015}%
  \BibitemOpen
  \bibfield  {author} {\bibinfo {author} {\bibfnamefont {S.}~\bibnamefont
  {Wang}}, \bibinfo {author} {\bibfnamefont {X.}~\bibnamefont {Wang}},\ and\
  \bibinfo {author} {\bibfnamefont {J.~H.}\ \bibnamefont {Warner}},\ }\bibfield
   {title} {\bibinfo {title} {All chemical vapor deposition growth of
  {M}o{S}$_2$:h-{BN} vertical van der {W}aals heterostructures},\ }\href
  {https://doi.org/10.1021/acsnano.5b00655} {\bibfield  {journal} {\bibinfo
  {journal} {ACS Nano}\ }\textbf {\bibinfo {volume} {9}},\ \bibinfo {pages}
  {5246} (\bibinfo {year} {2015})}\BibitemShut {NoStop}%
\bibitem [{\citenamefont {Ji}\ \emph {et~al.}(2015)\citenamefont {Ji},
  \citenamefont {Zhang}, \citenamefont {Zhang},\ and\ \citenamefont
  {Liu}}]{Ji2015}%
  \BibitemOpen
  \bibfield  {author} {\bibinfo {author} {\bibfnamefont {Q.}~\bibnamefont
  {Ji}}, \bibinfo {author} {\bibfnamefont {Y.}~\bibnamefont {Zhang}}, \bibinfo
  {author} {\bibfnamefont {Y.}~\bibnamefont {Zhang}},\ and\ \bibinfo {author}
  {\bibfnamefont {Z.}~\bibnamefont {Liu}},\ }\bibfield  {title} {\bibinfo
  {title} {Chemical vapour deposition of group-{VIB} metal dichalcogenide
  monolayers: engineered substrates from amorphous to single crystalline},\
  }\href {https://doi.org/10.1039/C4CS00258J} {\bibfield  {journal} {\bibinfo
  {journal} {Chem. Soc. Rev.}\ }\textbf {\bibinfo {volume} {44}},\ \bibinfo
  {pages} {2587} (\bibinfo {year} {2015})}\BibitemShut {NoStop}%
\bibitem [{\citenamefont {Sebastian}\ \emph {et~al.}(2021)\citenamefont
  {Sebastian}, \citenamefont {Pendurthi}, \citenamefont {Choudhury},
  \citenamefont {Redwing},\ and\ \citenamefont {Das}}]{Sebastian2021}%
  \BibitemOpen
  \bibfield  {author} {\bibinfo {author} {\bibfnamefont {A.}~\bibnamefont
  {Sebastian}}, \bibinfo {author} {\bibfnamefont {R.}~\bibnamefont
  {Pendurthi}}, \bibinfo {author} {\bibfnamefont {T.~H.}\ \bibnamefont
  {Choudhury}}, \bibinfo {author} {\bibfnamefont {J.~M.}\ \bibnamefont
  {Redwing}},\ and\ \bibinfo {author} {\bibfnamefont {S.}~\bibnamefont {Das}},\
  }\bibfield  {title} {\bibinfo {title} {Benchmarking monolayer {M}o{S}$_2$ and
  {WS}$_2$ field-effect transistors},\ }\href
  {https://doi.org/10.1038/s41467-020-20732-w} {\bibfield  {journal} {\bibinfo
  {journal} {Nat. Commun.}\ }\textbf {\bibinfo {volume} {12}},\ \bibinfo
  {pages} {693} (\bibinfo {year} {2021})}\BibitemShut {NoStop}%
\bibitem [{\citenamefont {Liu}\ \emph {et~al.}(2013)\citenamefont {Liu},
  \citenamefont {Kang}, \citenamefont {Sarkar}, \citenamefont {Khatami},
  \citenamefont {Jena},\ and\ \citenamefont {Banerjee}}]{Liu2013}%
  \BibitemOpen
  \bibfield  {author} {\bibinfo {author} {\bibfnamefont {W.}~\bibnamefont
  {Liu}}, \bibinfo {author} {\bibfnamefont {J.}~\bibnamefont {Kang}}, \bibinfo
  {author} {\bibfnamefont {D.}~\bibnamefont {Sarkar}}, \bibinfo {author}
  {\bibfnamefont {Y.}~\bibnamefont {Khatami}}, \bibinfo {author} {\bibfnamefont
  {D.}~\bibnamefont {Jena}},\ and\ \bibinfo {author} {\bibfnamefont
  {K.}~\bibnamefont {Banerjee}},\ }\bibfield  {title} {\bibinfo {title} {Role
  of metal contacts in designing high-performance monolayer n-type {WS}e$_2$
  field effect transistors},\ }\href {https://doi.org/10.1021/nl304777e}
  {\bibfield  {journal} {\bibinfo  {journal} {Nano Lett.}\ }\textbf {\bibinfo
  {volume} {13}},\ \bibinfo {pages} {1983} (\bibinfo {year}
  {2013})}\BibitemShut {NoStop}%
\bibitem [{\citenamefont {Podzorov}\ \emph {et~al.}(2004)\citenamefont
  {Podzorov}, \citenamefont {Gershenson}, \citenamefont {Kloc}, \citenamefont
  {Zeis},\ and\ \citenamefont {Bucher}}]{Podzorov2004}%
  \BibitemOpen
  \bibfield  {author} {\bibinfo {author} {\bibfnamefont {V.}~\bibnamefont
  {Podzorov}}, \bibinfo {author} {\bibfnamefont {M.~E.}\ \bibnamefont
  {Gershenson}}, \bibinfo {author} {\bibfnamefont {C.}~\bibnamefont {Kloc}},
  \bibinfo {author} {\bibfnamefont {R.}~\bibnamefont {Zeis}},\ and\ \bibinfo
  {author} {\bibfnamefont {E.}~\bibnamefont {Bucher}},\ }\bibfield  {title}
  {\bibinfo {title} {High-mobility field-effect transistors based on transition
  metal dichalcogenides},\ }\href {https://doi.org/10.1063/1.1723695}
  {\bibfield  {journal} {\bibinfo  {journal} {Appl. Phys. Lett.}\ }\textbf
  {\bibinfo {volume} {84}},\ \bibinfo {pages} {3301} (\bibinfo {year}
  {2004})}\BibitemShut {NoStop}%
\bibitem [{\citenamefont {Ryou}\ \emph {et~al.}(2016)\citenamefont {Ryou},
  \citenamefont {Kim}, \citenamefont {Santosh},\ and\ \citenamefont
  {Cho}}]{Ryou2016}%
  \BibitemOpen
  \bibfield  {author} {\bibinfo {author} {\bibfnamefont {J.}~\bibnamefont
  {Ryou}}, \bibinfo {author} {\bibfnamefont {Y.-S.}\ \bibnamefont {Kim}},
  \bibinfo {author} {\bibfnamefont {K.~C.}\ \bibnamefont {Santosh}},\ and\
  \bibinfo {author} {\bibfnamefont {K.}~\bibnamefont {Cho}},\ }\bibfield
  {title} {\bibinfo {title} {Monolayer {MoS}$_2$ bandgap modulation by
  dielectric environments and tunable bandgap transistors},\ }\href
  {https://doi.org/10.1038/srep29184} {\bibfield  {journal} {\bibinfo
  {journal} {Sci. Rep.}\ }\textbf {\bibinfo {volume} {6}},\ \bibinfo {pages}
  {29184} (\bibinfo {year} {2016})}\BibitemShut {NoStop}%
\bibitem [{\citenamefont {Hill}\ \emph {et~al.}(2016)\citenamefont {Hill},
  \citenamefont {Rigosi}, \citenamefont {Rim}, \citenamefont {Flynn},\ and\
  \citenamefont {Heinz}}]{Hill2016}%
  \BibitemOpen
  \bibfield  {author} {\bibinfo {author} {\bibfnamefont {H.~M.}\ \bibnamefont
  {Hill}}, \bibinfo {author} {\bibfnamefont {A.~F.}\ \bibnamefont {Rigosi}},
  \bibinfo {author} {\bibfnamefont {K.~T.}\ \bibnamefont {Rim}}, \bibinfo
  {author} {\bibfnamefont {G.~W.}\ \bibnamefont {Flynn}},\ and\ \bibinfo
  {author} {\bibfnamefont {T.~F.}\ \bibnamefont {Heinz}},\ }\bibfield  {title}
  {\bibinfo {title} {Band alignment in {M}o{S}$_2$/{WS}$_2$ transition metal
  dichalcogenide heterostructures probed by scanning tunneling microscopy and
  spectroscopy},\ }\href {https://doi.org/10.1021/acs.nanolett.6b01007}
  {\bibfield  {journal} {\bibinfo  {journal} {Nano Lett.}\ }\textbf {\bibinfo
  {volume} {16}},\ \bibinfo {pages} {4831} (\bibinfo {year}
  {2016})}\BibitemShut {NoStop}%
\bibitem [{\citenamefont {Qiu}\ \emph {et~al.}(2017)\citenamefont {Qiu},
  \citenamefont {da~Jornada},\ and\ \citenamefont {Louie}}]{Qiu2017}%
  \BibitemOpen
  \bibfield  {author} {\bibinfo {author} {\bibfnamefont {D.~Y.}\ \bibnamefont
  {Qiu}}, \bibinfo {author} {\bibfnamefont {F.~H.}\ \bibnamefont
  {da~Jornada}},\ and\ \bibinfo {author} {\bibfnamefont {S.~G.}\ \bibnamefont
  {Louie}},\ }\bibfield  {title} {\bibinfo {title} {Environmental screening
  effects in {2D} materials: Renormalization of the bandgap, electronic
  structure, and optical spectra of few-layer black phosphorus},\ }\href
  {https://doi.org/10.1021/acs.nanolett.7b01365} {\bibfield  {journal}
  {\bibinfo  {journal} {Nano Lett.}\ }\textbf {\bibinfo {volume} {17}},\
  \bibinfo {pages} {4706} (\bibinfo {year} {2017})}\BibitemShut {NoStop}%
\bibitem [{\citenamefont {Raja}\ \emph {et~al.}(2017)\citenamefont {Raja},
  \citenamefont {Chaves}, \citenamefont {Yu}, \citenamefont {Arefe},
  \citenamefont {Hill}, \citenamefont {Rigosi}, \citenamefont {Berkelbach},
  \citenamefont {Nagler}, \citenamefont {Schüller}, \citenamefont {Korn},
  \citenamefont {Nuckolls}, \citenamefont {Hone}, \citenamefont {Brus},
  \citenamefont {Heinz}, \citenamefont {Reichman},\ and\ \citenamefont
  {Chernikov}}]{Raja2017}%
  \BibitemOpen
  \bibfield  {author} {\bibinfo {author} {\bibfnamefont {A.}~\bibnamefont
  {Raja}}, \bibinfo {author} {\bibfnamefont {A.}~\bibnamefont {Chaves}},
  \bibinfo {author} {\bibfnamefont {J.}~\bibnamefont {Yu}}, \bibinfo {author}
  {\bibfnamefont {G.}~\bibnamefont {Arefe}}, \bibinfo {author} {\bibfnamefont
  {H.~M.}\ \bibnamefont {Hill}}, \bibinfo {author} {\bibfnamefont {A.~F.}\
  \bibnamefont {Rigosi}}, \bibinfo {author} {\bibfnamefont {T.~C.}\
  \bibnamefont {Berkelbach}}, \bibinfo {author} {\bibfnamefont
  {P.}~\bibnamefont {Nagler}}, \bibinfo {author} {\bibfnamefont
  {C.}~\bibnamefont {Schüller}}, \bibinfo {author} {\bibfnamefont
  {T.}~\bibnamefont {Korn}}, \bibinfo {author} {\bibfnamefont {C.}~\bibnamefont
  {Nuckolls}}, \bibinfo {author} {\bibfnamefont {J.}~\bibnamefont {Hone}},
  \bibinfo {author} {\bibfnamefont {L.~E.}\ \bibnamefont {Brus}}, \bibinfo
  {author} {\bibfnamefont {T.~F.}\ \bibnamefont {Heinz}}, \bibinfo {author}
  {\bibfnamefont {D.~R.}\ \bibnamefont {Reichman}},\ and\ \bibinfo {author}
  {\bibfnamefont {A.}~\bibnamefont {Chernikov}},\ }\bibfield  {title} {\bibinfo
  {title} {Coulomb engineering of the bandgap and excitons in two-dimensional
  materials},\ }\href {https://doi.org/10.1038/ncomms15251} {\bibfield
  {journal} {\bibinfo  {journal} {Nat. Commun.}\ }\textbf {\bibinfo {volume}
  {8}},\ \bibinfo {pages} {15251} (\bibinfo {year} {2017})}\BibitemShut
  {NoStop}%
\bibitem [{\citenamefont {Zheng}\ \emph {et~al.}(2016)\citenamefont {Zheng},
  \citenamefont {Huang}, \citenamefont {Chen}, \citenamefont {Zhao},
  \citenamefont {Eda}, \citenamefont {Spataru}, \citenamefont {Zhang},
  \citenamefont {Chang}, \citenamefont {Li}, \citenamefont {Chi}, \citenamefont
  {Quek},\ and\ \citenamefont {Wee}}]{Zheng2016}%
  \BibitemOpen
  \bibfield  {author} {\bibinfo {author} {\bibfnamefont {Y.~J.}\ \bibnamefont
  {Zheng}}, \bibinfo {author} {\bibfnamefont {Y.~L.}\ \bibnamefont {Huang}},
  \bibinfo {author} {\bibfnamefont {Y.}~\bibnamefont {Chen}}, \bibinfo {author}
  {\bibfnamefont {W.}~\bibnamefont {Zhao}}, \bibinfo {author} {\bibfnamefont
  {G.}~\bibnamefont {Eda}}, \bibinfo {author} {\bibfnamefont {C.~D.}\
  \bibnamefont {Spataru}}, \bibinfo {author} {\bibfnamefont {W.}~\bibnamefont
  {Zhang}}, \bibinfo {author} {\bibfnamefont {Y.-H.}\ \bibnamefont {Chang}},
  \bibinfo {author} {\bibfnamefont {L.-J.}\ \bibnamefont {Li}}, \bibinfo
  {author} {\bibfnamefont {D.}~\bibnamefont {Chi}}, \bibinfo {author}
  {\bibfnamefont {S.~Y.}\ \bibnamefont {Quek}},\ and\ \bibinfo {author}
  {\bibfnamefont {A.~T.~S.}\ \bibnamefont {Wee}},\ }\bibfield  {title}
  {\bibinfo {title} {Heterointerface screening effects between organic
  monolayers and monolayer transition metal dichalcogenides},\ }\href
  {https://doi.org/10.1021/acsnano.5b07314} {\bibfield  {journal} {\bibinfo
  {journal} {ACS Nano}\ }\textbf {\bibinfo {volume} {10}},\ \bibinfo {pages}
  {2476} (\bibinfo {year} {2016})}\BibitemShut {NoStop}%
\bibitem [{\citenamefont {Cho}\ and\ \citenamefont
  {Berkelbach}(2018)}]{Cho2018}%
  \BibitemOpen
  \bibfield  {author} {\bibinfo {author} {\bibfnamefont {Y.}~\bibnamefont
  {Cho}}\ and\ \bibinfo {author} {\bibfnamefont {T.~C.}\ \bibnamefont
  {Berkelbach}},\ }\bibfield  {title} {\bibinfo {title} {Environmentally
  sensitive theory of electronic and optical transitions in atomically thin
  semiconductors},\ }\href@noop {} {\bibfield  {journal} {\bibinfo  {journal}
  {Phys. Rev. B}\ }\textbf {\bibinfo {volume} {97}},\ \bibinfo {pages}
  {041409(R)} (\bibinfo {year} {2018})}\BibitemShut {NoStop}%
\bibitem [{\citenamefont {Zibouche}\ \emph {et~al.}(2021)\citenamefont
  {Zibouche}, \citenamefont {Schlipf},\ and\ \citenamefont
  {Giustino}}]{Zibouche2021}%
  \BibitemOpen
  \bibfield  {author} {\bibinfo {author} {\bibfnamefont {N.}~\bibnamefont
  {Zibouche}}, \bibinfo {author} {\bibfnamefont {M.}~\bibnamefont {Schlipf}},\
  and\ \bibinfo {author} {\bibfnamefont {F.}~\bibnamefont {Giustino}},\
  }\bibfield  {title} {\bibinfo {title} {{GW} band structure of monolayer
  {MoS$_2$} using the {SternheimerGW} method and effect of dielectric
  environment},\ }\href {https://doi.org/10.1103/PhysRevB.103.125401}
  {\bibfield  {journal} {\bibinfo  {journal} {Phys. Rev. B}\ }\textbf {\bibinfo
  {volume} {103}},\ \bibinfo {pages} {125401} (\bibinfo {year}
  {2021})}\BibitemShut {NoStop}%
\bibitem [{\citenamefont {Lattyak}\ \emph {et~al.}(2022)\citenamefont
  {Lattyak}, \citenamefont {Gehrke},\ and\ \citenamefont
  {Vehse}}]{Lattyak2022}%
  \BibitemOpen
  \bibfield  {author} {\bibinfo {author} {\bibfnamefont {C.}~\bibnamefont
  {Lattyak}}, \bibinfo {author} {\bibfnamefont {K.}~\bibnamefont {Gehrke}},\
  and\ \bibinfo {author} {\bibfnamefont {M.}~\bibnamefont {Vehse}},\ }\bibfield
   {title} {\bibinfo {title} {Layer-thickness-dependent work function of
  {M}o{S}$_2$ on metal and metal oxide substrates},\ }\href
  {https://doi.org/10.1021/acs.jpcc.2c03268} {\bibfield  {journal} {\bibinfo
  {journal} {J. Phys. Chem. C}\ }\textbf {\bibinfo {volume} {126}},\ \bibinfo
  {pages} {13929} (\bibinfo {year} {2022})}\BibitemShut {NoStop}%
\bibitem [{\citenamefont {Panasci}\ \emph {et~al.}(2021)\citenamefont
  {Panasci}, \citenamefont {Schilirò}, \citenamefont {Greco}, \citenamefont
  {Cannas}, \citenamefont {Gelardi}, \citenamefont {Agnello}, \citenamefont
  {Roccaforte},\ and\ \citenamefont {Giannazzo}}]{Panasci2021}%
  \BibitemOpen
  \bibfield  {author} {\bibinfo {author} {\bibfnamefont {S.~E.}\ \bibnamefont
  {Panasci}}, \bibinfo {author} {\bibfnamefont {E.}~\bibnamefont {Schilirò}},
  \bibinfo {author} {\bibfnamefont {G.}~\bibnamefont {Greco}}, \bibinfo
  {author} {\bibfnamefont {M.}~\bibnamefont {Cannas}}, \bibinfo {author}
  {\bibfnamefont {F.~M.}\ \bibnamefont {Gelardi}}, \bibinfo {author}
  {\bibfnamefont {S.}~\bibnamefont {Agnello}}, \bibinfo {author} {\bibfnamefont
  {F.}~\bibnamefont {Roccaforte}},\ and\ \bibinfo {author} {\bibfnamefont
  {F.}~\bibnamefont {Giannazzo}},\ }\bibfield  {title} {\bibinfo {title}
  {Strain, doping, and electronic transport of large area monolayer {M}o{S}$_2$
  exfoliated on gold and transferred to an insulating substrate},\ }\href
  {https://doi.org/10.1021/acsami.1c05185} {\bibfield  {journal} {\bibinfo
  {journal} {ACS Appl. Mater. Interfaces}\ }\textbf {\bibinfo {volume} {13}},\
  \bibinfo {pages} {31248} (\bibinfo {year} {2021})}\BibitemShut {NoStop}%
\bibitem [{\citenamefont {S{\o}rensen}\ \emph {et~al.}(2014)\citenamefont
  {S{\o}rensen}, \citenamefont {F\"{u}chtbauer}, \citenamefont {Tuxen},
  \citenamefont {Walton},\ and\ \citenamefont {Lauritsen}}]{Soerensen2014}%
  \BibitemOpen
  \bibfield  {author} {\bibinfo {author} {\bibfnamefont {S.~G.}\ \bibnamefont
  {S{\o}rensen}}, \bibinfo {author} {\bibfnamefont {H.~G.}\ \bibnamefont
  {F\"{u}chtbauer}}, \bibinfo {author} {\bibfnamefont {A.~K.}\ \bibnamefont
  {Tuxen}}, \bibinfo {author} {\bibfnamefont {A.~S.}\ \bibnamefont {Walton}},\
  and\ \bibinfo {author} {\bibfnamefont {J.~V.}\ \bibnamefont {Lauritsen}},\
  }\bibfield  {title} {\bibinfo {title} {Structure and electronic properties of
  in situ synthesized single-layer {M}o{S}$_2$ on a gold surface},\ }\href
  {https://doi.org/10.1021/nn502812n} {\bibfield  {journal} {\bibinfo
  {journal} {ACS Nano}\ }\textbf {\bibinfo {volume} {8}},\ \bibinfo {pages}
  {6788} (\bibinfo {year} {2014})}\BibitemShut {NoStop}%
\bibitem [{\citenamefont {Liu}\ \emph {et~al.}(2016)\citenamefont {Liu},
  \citenamefont {Balla}, \citenamefont {Bergeron}, \citenamefont {Campbell},
  \citenamefont {Bedzyk},\ and\ \citenamefont {Hersam}}]{Liu2016}%
  \BibitemOpen
  \bibfield  {author} {\bibinfo {author} {\bibfnamefont {X.}~\bibnamefont
  {Liu}}, \bibinfo {author} {\bibfnamefont {I.}~\bibnamefont {Balla}}, \bibinfo
  {author} {\bibfnamefont {H.}~\bibnamefont {Bergeron}}, \bibinfo {author}
  {\bibfnamefont {G.~P.}\ \bibnamefont {Campbell}}, \bibinfo {author}
  {\bibfnamefont {M.~J.}\ \bibnamefont {Bedzyk}},\ and\ \bibinfo {author}
  {\bibfnamefont {M.~C.}\ \bibnamefont {Hersam}},\ }\bibfield  {title}
  {\bibinfo {title} {Rotationally commensurate growth of {M}o{S}$_2$ on
  epitaxial graphene},\ }\href {https://doi.org/10.1021/acsnano.5b06398}
  {\bibfield  {journal} {\bibinfo  {journal} {ACS Nano}\ }\textbf {\bibinfo
  {volume} {10}},\ \bibinfo {pages} {1067} (\bibinfo {year}
  {2016})}\BibitemShut {NoStop}%
\bibitem [{\citenamefont {Huang}\ \emph {et~al.}(2015)\citenamefont {Huang},
  \citenamefont {Chen}, \citenamefont {Zhang}, \citenamefont {Quek},
  \citenamefont {Chen}, \citenamefont {Li}, \citenamefont {Hsu}, \citenamefont
  {Chang}, \citenamefont {Zheng}, \citenamefont {Chen},\ and\ \citenamefont
  {Wee}}]{Huang2015}%
  \BibitemOpen
  \bibfield  {author} {\bibinfo {author} {\bibfnamefont {Y.~L.}\ \bibnamefont
  {Huang}}, \bibinfo {author} {\bibfnamefont {Y.}~\bibnamefont {Chen}},
  \bibinfo {author} {\bibfnamefont {W.}~\bibnamefont {Zhang}}, \bibinfo
  {author} {\bibfnamefont {S.~Y.}\ \bibnamefont {Quek}}, \bibinfo {author}
  {\bibfnamefont {C.-H.}\ \bibnamefont {Chen}}, \bibinfo {author}
  {\bibfnamefont {L.-J.}\ \bibnamefont {Li}}, \bibinfo {author} {\bibfnamefont
  {W.-T.}\ \bibnamefont {Hsu}}, \bibinfo {author} {\bibfnamefont {W.-H.}\
  \bibnamefont {Chang}}, \bibinfo {author} {\bibfnamefont {Y.~J.}\ \bibnamefont
  {Zheng}}, \bibinfo {author} {\bibfnamefont {W.}~\bibnamefont {Chen}},\ and\
  \bibinfo {author} {\bibfnamefont {A.~T.~S.}\ \bibnamefont {Wee}},\ }\bibfield
   {title} {\bibinfo {title} {Bandgap tunability at single-layer molybdenum
  disulphide grain boundaries},\ }\href {https://doi.org/10.1038/ncomms7298}
  {\bibfield  {journal} {\bibinfo  {journal} {Nat. Commun.}\ }\textbf {\bibinfo
  {volume} {6}},\ \bibinfo {pages} {6298} (\bibinfo {year} {2015})}\BibitemShut
  {NoStop}%
\bibitem [{\citenamefont {Chiu}\ \emph {et~al.}(2015)\citenamefont {Chiu},
  \citenamefont {Zhang}, \citenamefont {Shiu}, \citenamefont {Chuu},
  \citenamefont {Chen}, \citenamefont {Chang}, \citenamefont {Chen},
  \citenamefont {Chou}, \citenamefont {Shih},\ and\ \citenamefont
  {Li}}]{Chiu2015}%
  \BibitemOpen
  \bibfield  {author} {\bibinfo {author} {\bibfnamefont {M.-H.}\ \bibnamefont
  {Chiu}}, \bibinfo {author} {\bibfnamefont {C.}~\bibnamefont {Zhang}},
  \bibinfo {author} {\bibfnamefont {H.-W.}\ \bibnamefont {Shiu}}, \bibinfo
  {author} {\bibfnamefont {C.-P.}\ \bibnamefont {Chuu}}, \bibinfo {author}
  {\bibfnamefont {C.-H.}\ \bibnamefont {Chen}}, \bibinfo {author}
  {\bibfnamefont {C.-Y.~S.}\ \bibnamefont {Chang}}, \bibinfo {author}
  {\bibfnamefont {C.-H.}\ \bibnamefont {Chen}}, \bibinfo {author}
  {\bibfnamefont {M.-Y.}\ \bibnamefont {Chou}}, \bibinfo {author}
  {\bibfnamefont {C.-K.}\ \bibnamefont {Shih}},\ and\ \bibinfo {author}
  {\bibfnamefont {L.-J.}\ \bibnamefont {Li}},\ }\bibfield  {title} {\bibinfo
  {title} {Determination of band alignment in the single-layer
  {M}o{S}$_2$/{WS}e$_2$ heterojunction},\ }\href
  {https://doi.org/10.1038/ncomms8666} {\bibfield  {journal} {\bibinfo
  {journal} {Nat. Commun.}\ }\textbf {\bibinfo {volume} {6}},\ \bibinfo {pages}
  {7666} (\bibinfo {year} {2015})}\BibitemShut {NoStop}%
\bibitem [{\citenamefont {Zhang}\ \emph {et~al.}(2014)\citenamefont {Zhang},
  \citenamefont {Johnson}, \citenamefont {Hsu}, \citenamefont {Li},\ and\
  \citenamefont {Shih}}]{Zhang2014}%
  \BibitemOpen
  \bibfield  {author} {\bibinfo {author} {\bibfnamefont {C.}~\bibnamefont
  {Zhang}}, \bibinfo {author} {\bibfnamefont {A.}~\bibnamefont {Johnson}},
  \bibinfo {author} {\bibfnamefont {C.-L.}\ \bibnamefont {Hsu}}, \bibinfo
  {author} {\bibfnamefont {L.-J.}\ \bibnamefont {Li}},\ and\ \bibinfo {author}
  {\bibfnamefont {C.-K.}\ \bibnamefont {Shih}},\ }\bibfield  {title} {\bibinfo
  {title} {Direct imaging of band profile in single layer {M}o{S}$_2$ on
  graphite: Quasiparticle energy gap, metallic edge states, and edge band
  bending},\ }\href {https://doi.org/10.1021/nl501133c} {\bibfield  {journal}
  {\bibinfo  {journal} {Nano Lett.}\ }\textbf {\bibinfo {volume} {14}},\
  \bibinfo {pages} {2443} (\bibinfo {year} {2014})}\BibitemShut {NoStop}%
\bibitem [{\citenamefont {Li}\ \emph {et~al.}(2014)\citenamefont {Li},
  \citenamefont {Wu}, \citenamefont {Yin},\ and\ \citenamefont
  {Zhang}}]{Li2014}%
  \BibitemOpen
  \bibfield  {author} {\bibinfo {author} {\bibfnamefont {H.}~\bibnamefont
  {Li}}, \bibinfo {author} {\bibfnamefont {J.}~\bibnamefont {Wu}}, \bibinfo
  {author} {\bibfnamefont {Z.}~\bibnamefont {Yin}},\ and\ \bibinfo {author}
  {\bibfnamefont {H.}~\bibnamefont {Zhang}},\ }\bibfield  {title} {\bibinfo
  {title} {Preparation and applications of mechanically exfoliated single-layer
  and multi layer {M}o{S}$_2$ and {WS}e$_2$ nanosheets},\ }\href@noop {}
  {\bibfield  {journal} {\bibinfo  {journal} {Acc. Chem. Res.}\ }\textbf
  {\bibinfo {volume} {47}},\ \bibinfo {pages} {1067} (\bibinfo {year}
  {2014})}\BibitemShut {NoStop}%
\bibitem [{\citenamefont {Lu}\ \emph {et~al.}(2014)\citenamefont {Lu},
  \citenamefont {Li}, \citenamefont {Mao}, \citenamefont {Wang},\ and\
  \citenamefont {Andrei}}]{Lu2014}%
  \BibitemOpen
  \bibfield  {author} {\bibinfo {author} {\bibfnamefont {C.-P.}\ \bibnamefont
  {Lu}}, \bibinfo {author} {\bibfnamefont {G.}~\bibnamefont {Li}}, \bibinfo
  {author} {\bibfnamefont {J.}~\bibnamefont {Mao}}, \bibinfo {author}
  {\bibfnamefont {L.-M.}\ \bibnamefont {Wang}},\ and\ \bibinfo {author}
  {\bibfnamefont {E.~Y.}\ \bibnamefont {Andrei}},\ }\bibfield  {title}
  {\bibinfo {title} {Bandgap, mid-gap states, and gating effects in
  {M}o{S}$_2$},\ }\href@noop {} {\bibfield  {journal} {\bibinfo  {journal}
  {Nano Lett.}\ }\textbf {\bibinfo {volume} {14}},\ \bibinfo {pages} {4628}
  (\bibinfo {year} {2014})}\BibitemShut {NoStop}%
\bibitem [{\citenamefont {Chow}\ \emph {et~al.}(2015)\citenamefont {Chow},
  \citenamefont {Singh}, \citenamefont {Viana}, \citenamefont {Gao},
  \citenamefont {Luo}, \citenamefont {Li}, \citenamefont {Lin}, \citenamefont
  {Elias}, \citenamefont {Shi}, \citenamefont {Wang}, \citenamefont
  {Terrones},\ and\ \citenamefont {Koratkar}}]{Chow2015}%
  \BibitemOpen
  \bibfield  {author} {\bibinfo {author} {\bibfnamefont {P.~K.}\ \bibnamefont
  {Chow}}, \bibinfo {author} {\bibfnamefont {E.}~\bibnamefont {Singh}},
  \bibinfo {author} {\bibfnamefont {B.~C.}\ \bibnamefont {Viana}}, \bibinfo
  {author} {\bibfnamefont {J.}~\bibnamefont {Gao}}, \bibinfo {author}
  {\bibfnamefont {J.}~\bibnamefont {Luo}}, \bibinfo {author} {\bibfnamefont
  {J.}~\bibnamefont {Li}}, \bibinfo {author} {\bibfnamefont {Z.}~\bibnamefont
  {Lin}}, \bibinfo {author} {\bibfnamefont {A.~L.}\ \bibnamefont {Elias}},
  \bibinfo {author} {\bibfnamefont {Y.}~\bibnamefont {Shi}}, \bibinfo {author}
  {\bibfnamefont {Z.}~\bibnamefont {Wang}}, \bibinfo {author} {\bibfnamefont
  {M.}~\bibnamefont {Terrones}},\ and\ \bibinfo {author} {\bibfnamefont
  {N.}~\bibnamefont {Koratkar}},\ }\bibfield  {title} {\bibinfo {title}
  {Wetting of mono and few-layered {WS}$_2$ and {M}o{S}$_2$ films supported on
  {S}i/{S}i{O}$_2$ substrates},\ }\href@noop {} {\bibfield  {journal} {\bibinfo
   {journal} {ACS Nano}\ }\textbf {\bibinfo {volume} {9}},\ \bibinfo {pages}
  {3023} (\bibinfo {year} {2015})}\BibitemShut {NoStop}%
\bibitem [{\citenamefont {Han}\ \emph {et~al.}(2019)\citenamefont {Han},
  \citenamefont {Lin}, \citenamefont {Liu}, \citenamefont {Wang},\ and\
  \citenamefont {Pan}}]{Han2019}%
  \BibitemOpen
  \bibfield  {author} {\bibinfo {author} {\bibfnamefont {X.}~\bibnamefont
  {Han}}, \bibinfo {author} {\bibfnamefont {J.}~\bibnamefont {Lin}}, \bibinfo
  {author} {\bibfnamefont {J.}~\bibnamefont {Liu}}, \bibinfo {author}
  {\bibfnamefont {N.}~\bibnamefont {Wang}},\ and\ \bibinfo {author}
  {\bibfnamefont {D.}~\bibnamefont {Pan}},\ }\bibfield  {title} {\bibinfo
  {title} {Effects of hexagonal boron nitride encapsulation on the electronic
  structure of few-layer {M}o{S}$_2$},\ }\href
  {https://doi.org/10.1021/acs.jpcc.9b02549} {\bibfield  {journal} {\bibinfo
  {journal} {J. Phys. Chem. C}\ }\textbf {\bibinfo {volume} {123}},\ \bibinfo
  {pages} {14797} (\bibinfo {year} {2019})}\BibitemShut {NoStop}%
\bibitem [{\citenamefont {Zhang}\ \emph {et~al.}(2019)\citenamefont {Zhang},
  \citenamefont {Zhang}, \citenamefont {Wang}, \citenamefont {Schulman},
  \citenamefont {Zhang}, \citenamefont {Bansal}, \citenamefont {Alem},
  \citenamefont {Das}, \citenamefont {Crespi},\ and\ \citenamefont
  {Terrones}}]{Zhang2019}%
  \BibitemOpen
  \bibfield  {author} {\bibinfo {author} {\bibfnamefont {X.}~\bibnamefont
  {Zhang}}, \bibinfo {author} {\bibfnamefont {F.}~\bibnamefont {Zhang}},
  \bibinfo {author} {\bibfnamefont {Y.}~\bibnamefont {Wang}}, \bibinfo {author}
  {\bibfnamefont {D.~S.}\ \bibnamefont {Schulman}}, \bibinfo {author}
  {\bibfnamefont {T.}~\bibnamefont {Zhang}}, \bibinfo {author} {\bibfnamefont
  {A.}~\bibnamefont {Bansal}}, \bibinfo {author} {\bibfnamefont
  {N.}~\bibnamefont {Alem}}, \bibinfo {author} {\bibfnamefont {S.}~\bibnamefont
  {Das}}, \bibinfo {author} {\bibfnamefont {V.~H.}\ \bibnamefont {Crespi}},\
  and\ \bibinfo {author} {\bibfnamefont {M.}~\bibnamefont {Terrones}},\
  }\bibfield  {title} {\bibinfo {title} {Defect-controlled nucleation and
  orientation of {WS}e$_2$ on h{BN}: A route to single-crystal epitaxial
  monolayers},\ }\href@noop {} {\bibfield  {journal} {\bibinfo  {journal} {ACS
  Nano}\ }\textbf {\bibinfo {volume} {13}},\ \bibinfo {pages} {3341} (\bibinfo
  {year} {2019})}\BibitemShut {NoStop}%
\bibitem [{\citenamefont {Penn}(1962)}]{Penn1962}%
  \BibitemOpen
  \bibfield  {author} {\bibinfo {author} {\bibfnamefont {D.~R.}\ \bibnamefont
  {Penn}},\ }\bibfield  {title} {\bibinfo {title} {Wave-number-dependent
  dielectric function of semiconductors},\ }\href
  {https://doi.org/10.1103/PhysRev.128.2093} {\bibfield  {journal} {\bibinfo
  {journal} {Phys. Rev.}\ }\textbf {\bibinfo {volume} {128}},\ \bibinfo {pages}
  {2093} (\bibinfo {year} {1962})}\BibitemShut {NoStop}%
\bibitem [{\citenamefont {Klein}\ \emph {et~al.}(2019)\citenamefont {Klein},
  \citenamefont {Kerelsky}, \citenamefont {Lorke}, \citenamefont {Florian},
  \citenamefont {Sigger}, \citenamefont {Kiemle}, \citenamefont {Reuter},
  \citenamefont {Taniguchi}, \citenamefont {Watanabe}, \citenamefont {Finley},
  \citenamefont {Pasupathy}, \citenamefont {Holleitner}, \citenamefont {Ross},\
  and\ \citenamefont {Wurstbauer}}]{Klein2019}%
  \BibitemOpen
  \bibfield  {author} {\bibinfo {author} {\bibfnamefont {J.}~\bibnamefont
  {Klein}}, \bibinfo {author} {\bibfnamefont {A.}~\bibnamefont {Kerelsky}},
  \bibinfo {author} {\bibfnamefont {M.}~\bibnamefont {Lorke}}, \bibinfo
  {author} {\bibfnamefont {M.}~\bibnamefont {Florian}}, \bibinfo {author}
  {\bibfnamefont {F.}~\bibnamefont {Sigger}}, \bibinfo {author} {\bibfnamefont
  {J.}~\bibnamefont {Kiemle}}, \bibinfo {author} {\bibfnamefont {M.~C.}\
  \bibnamefont {Reuter}}, \bibinfo {author} {\bibfnamefont {T.}~\bibnamefont
  {Taniguchi}}, \bibinfo {author} {\bibfnamefont {K.}~\bibnamefont {Watanabe}},
  \bibinfo {author} {\bibfnamefont {J.~J.}\ \bibnamefont {Finley}}, \bibinfo
  {author} {\bibfnamefont {A.~N.}\ \bibnamefont {Pasupathy}}, \bibinfo {author}
  {\bibfnamefont {A.~W.}\ \bibnamefont {Holleitner}}, \bibinfo {author}
  {\bibfnamefont {F.~M.}\ \bibnamefont {Ross}},\ and\ \bibinfo {author}
  {\bibfnamefont {U.}~\bibnamefont {Wurstbauer}},\ }\bibfield  {title}
  {\bibinfo {title} {Impact of substrate induced band tail states on the
  electronic and optical properties of {M}o{S}$_2$},\ }\href
  {https://doi.org/10.1063/1.5131270} {\bibfield  {journal} {\bibinfo
  {journal} {Appl. Phys. Lett.}\ }\textbf {\bibinfo {volume} {115}},\ \bibinfo
  {pages} {261603} (\bibinfo {year} {2019})}\BibitemShut {NoStop}%
\bibitem [{\citenamefont {K{\"u}mmel}\ and\ \citenamefont
  {Kronik}(2008)}]{Kuemmel2008}%
  \BibitemOpen
  \bibfield  {author} {\bibinfo {author} {\bibfnamefont {S.}~\bibnamefont
  {K{\"u}mmel}}\ and\ \bibinfo {author} {\bibfnamefont {L.}~\bibnamefont
  {Kronik}},\ }\bibfield  {title} {\bibinfo {title} {Orbital-dependent density
  functionals: Theory and applications},\ }\href
  {https://link.aps.org/doi/10.1103/RevModPhys.80.3} {\bibfield  {journal}
  {\bibinfo  {journal} {Rev. Mod. Phys.}\ }\textbf {\bibinfo {volume} {80}},\
  \bibinfo {pages} {3} (\bibinfo {year} {2008})}\BibitemShut {NoStop}%
\bibitem [{\citenamefont {Perdew}\ and\ \citenamefont
  {Levy}(1983)}]{Perdew1983}%
  \BibitemOpen
  \bibfield  {author} {\bibinfo {author} {\bibfnamefont {J.~P.}\ \bibnamefont
  {Perdew}}\ and\ \bibinfo {author} {\bibfnamefont {M.}~\bibnamefont {Levy}},\
  }\bibfield  {title} {\bibinfo {title} {Physical content of the exact
  kohn-sham orbital energies: Band gaps and derivative discontinuities},\
  }\href {https://link.aps.org/doi/10.1103/PhysRevLett.51.1884} {\bibfield
  {journal} {\bibinfo  {journal} {Phys. Rev. Lett.}\ }\textbf {\bibinfo
  {volume} {51}},\ \bibinfo {pages} {1884} (\bibinfo {year}
  {1983})}\BibitemShut {NoStop}%
\bibitem [{\citenamefont {Yang}\ \emph {et~al.}(2012)\citenamefont {Yang},
  \citenamefont {Cohen},\ and\ \citenamefont {Mori-S{\'a}nchez}}]{Yang2012}%
  \BibitemOpen
  \bibfield  {author} {\bibinfo {author} {\bibfnamefont {W.}~\bibnamefont
  {Yang}}, \bibinfo {author} {\bibfnamefont {A.~J.}\ \bibnamefont {Cohen}},\
  and\ \bibinfo {author} {\bibfnamefont {P.}~\bibnamefont {Mori-S{\'a}nchez}},\
  }\bibfield  {title} {\bibinfo {title} {Derivative discontinuity, bandgap and
  lowest unoccupied molecular orbital in density functional theory},\ }\href
  {https://doi.org/10.1063/1.3702391} {\bibfield  {journal} {\bibinfo
  {journal} {J. Chem. Phys.}\ }\textbf {\bibinfo {volume} {136}},\ \bibinfo
  {pages} {204111} (\bibinfo {year} {2012})}\BibitemShut {NoStop}%
\bibitem [{\citenamefont {Hedin}(1965)}]{Hedin1965}%
  \BibitemOpen
  \bibfield  {author} {\bibinfo {author} {\bibfnamefont {L.}~\bibnamefont
  {Hedin}},\ }\bibfield  {title} {\bibinfo {title} {New method for calculating
  the one-particle {G}reen's function with application to the electron-gas
  problem},\ }\href {https://doi.org/10.1103/PhysRev.139.A796} {\bibfield
  {journal} {\bibinfo  {journal} {Phys. Rev.}\ }\textbf {\bibinfo {volume}
  {139}},\ \bibinfo {pages} {A796} (\bibinfo {year} {1965})}\BibitemShut
  {NoStop}%
\bibitem [{\citenamefont {Strinati}\ \emph {et~al.}(1982)\citenamefont
  {Strinati}, \citenamefont {Mattausch},\ and\ \citenamefont
  {Hanke}}]{Strinati1982}%
  \BibitemOpen
  \bibfield  {author} {\bibinfo {author} {\bibfnamefont {G.}~\bibnamefont
  {Strinati}}, \bibinfo {author} {\bibfnamefont {H.~J.}\ \bibnamefont
  {Mattausch}},\ and\ \bibinfo {author} {\bibfnamefont {W.}~\bibnamefont
  {Hanke}},\ }\bibfield  {title} {\bibinfo {title} {Dynamical aspects of
  correlation corrections in a covalent crystal},\ }\href
  {https://doi.org/10.1103/PhysRevB.25.2867} {\bibfield  {journal} {\bibinfo
  {journal} {Phys. Rev. B}\ }\textbf {\bibinfo {volume} {25}},\ \bibinfo
  {pages} {2867} (\bibinfo {year} {1982})}\BibitemShut {NoStop}%
\bibitem [{\citenamefont {Hybertsen}\ and\ \citenamefont
  {Louie}(1986)}]{Hybertsen1986}%
  \BibitemOpen
  \bibfield  {author} {\bibinfo {author} {\bibfnamefont {M.~S.}\ \bibnamefont
  {Hybertsen}}\ and\ \bibinfo {author} {\bibfnamefont {S.~G.}\ \bibnamefont
  {Louie}},\ }\bibfield  {title} {\bibinfo {title} {Electron correlation in
  semiconductors and insulators: Band gaps and quasiparticle energies},\ }\href
  {https://doi.org/10.1103/PhysRevB.34.5390} {\bibfield  {journal} {\bibinfo
  {journal} {Phys. Rev. B}\ }\textbf {\bibinfo {volume} {34}},\ \bibinfo
  {pages} {5390} (\bibinfo {year} {1986})}\BibitemShut {NoStop}%
\bibitem [{\citenamefont {Rohlfing}\ and\ \citenamefont
  {Louie}(2000)}]{Rohlfing2000}%
  \BibitemOpen
  \bibfield  {author} {\bibinfo {author} {\bibfnamefont {M.}~\bibnamefont
  {Rohlfing}}\ and\ \bibinfo {author} {\bibfnamefont {S.~G.}\ \bibnamefont
  {Louie}},\ }\bibfield  {title} {\bibinfo {title} {Electron-hole excitations
  and optical spectra from first principles},\ }\href
  {https://doi.org/10.1103/PhysRevB.62.4927} {\bibfield  {journal} {\bibinfo
  {journal} {Phys. Rev. B}\ }\textbf {\bibinfo {volume} {62}},\ \bibinfo
  {pages} {4927} (\bibinfo {year} {2000})}\BibitemShut {NoStop}%
\bibitem [{\citenamefont {Shi}\ \emph {et~al.}(2013)\citenamefont {Shi},
  \citenamefont {Pan}, \citenamefont {Zhang},\ and\ \citenamefont
  {Yakobson}}]{Shi2013}%
  \BibitemOpen
  \bibfield  {author} {\bibinfo {author} {\bibfnamefont {H.}~\bibnamefont
  {Shi}}, \bibinfo {author} {\bibfnamefont {H.}~\bibnamefont {Pan}}, \bibinfo
  {author} {\bibfnamefont {Y.-W.}\ \bibnamefont {Zhang}},\ and\ \bibinfo
  {author} {\bibfnamefont {B.~I.}\ \bibnamefont {Yakobson}},\ }\bibfield
  {title} {\bibinfo {title} {Quasiparticle band structures and optical
  properties of strained monolayer {MoS}$_2$ and {WS}$_2$},\ }\href
  {https://doi.org/10.1103/PhysRevB.87.155304} {\bibfield  {journal} {\bibinfo
  {journal} {Phys. Rev. B}\ }\textbf {\bibinfo {volume} {87}},\ \bibinfo
  {pages} {155304} (\bibinfo {year} {2013})}\BibitemShut {NoStop}%
\bibitem [{\citenamefont {Qiu}\ \emph {et~al.}(2013)\citenamefont {Qiu},
  \citenamefont {da~Jornada},\ and\ \citenamefont {Louie}}]{Qiu2013}%
  \BibitemOpen
  \bibfield  {author} {\bibinfo {author} {\bibfnamefont {D.~Y.}\ \bibnamefont
  {Qiu}}, \bibinfo {author} {\bibfnamefont {F.~H.}\ \bibnamefont
  {da~Jornada}},\ and\ \bibinfo {author} {\bibfnamefont {S.~G.}\ \bibnamefont
  {Louie}},\ }\bibfield  {title} {\bibinfo {title} {Optical spectrum of
  {M}o{S}$_2$: Many-body effects and diversity of exciton states},\ }\href
  {https://doi.org/10.1103/PhysRevLett.111.216805} {\bibfield  {journal}
  {\bibinfo  {journal} {Phys. Rev. Lett.}\ }\textbf {\bibinfo {volume} {111}},\
  \bibinfo {pages} {216805} (\bibinfo {year} {2013})}\BibitemShut {NoStop}%
\bibitem [{\citenamefont {Ramasubramaniam}(2012)}]{Ramasubramaniam2012}%
  \BibitemOpen
  \bibfield  {author} {\bibinfo {author} {\bibfnamefont {A.}~\bibnamefont
  {Ramasubramaniam}},\ }\bibfield  {title} {\bibinfo {title} {Large excitonic
  effects in monolayers of molybdenum and tungsten dichalcogenides},\ }\href
  {https://doi.org/10.1103/PhysRevB.86.115409} {\bibfield  {journal} {\bibinfo
  {journal} {Phys. Rev. B}\ }\textbf {\bibinfo {volume} {86}},\ \bibinfo
  {pages} {115409} (\bibinfo {year} {2012})}\BibitemShut {NoStop}%
\bibitem [{\citenamefont {Cheiwchanchamnangij}\ and\ \citenamefont
  {Lambrecht}(2012)}]{Cheiwchanchamnangij2012}%
  \BibitemOpen
  \bibfield  {author} {\bibinfo {author} {\bibfnamefont {T.}~\bibnamefont
  {Cheiwchanchamnangij}}\ and\ \bibinfo {author} {\bibfnamefont {W.~R.~L.}\
  \bibnamefont {Lambrecht}},\ }\bibfield  {title} {\bibinfo {title}
  {Quasiparticle band structure calculation of monolayer, bilayer, and bulk
  {M}o{S}$_2$},\ }\href {https://doi.org/10.1103/PhysRevB.85.205302} {\bibfield
   {journal} {\bibinfo  {journal} {Phys. Rev. B}\ }\textbf {\bibinfo {volume}
  {85}},\ \bibinfo {pages} {205302} (\bibinfo {year} {2012})}\BibitemShut
  {NoStop}%
\bibitem [{\citenamefont {Ugeda}\ \emph {et~al.}(2014)\citenamefont {Ugeda},
  \citenamefont {Bradley}, \citenamefont {Shi}, \citenamefont {da~Jornada},
  \citenamefont {Zhang}, \citenamefont {Qiu}, \citenamefont {Ruan},
  \citenamefont {Mo}, \citenamefont {Hussain}, \citenamefont {Shen},
  \citenamefont {Wang}, \citenamefont {Louie},\ and\ \citenamefont
  {Crommie}}]{Ugeda2014}%
  \BibitemOpen
  \bibfield  {author} {\bibinfo {author} {\bibfnamefont {M.~M.}\ \bibnamefont
  {Ugeda}}, \bibinfo {author} {\bibfnamefont {A.~J.}\ \bibnamefont {Bradley}},
  \bibinfo {author} {\bibfnamefont {S.-F.}\ \bibnamefont {Shi}}, \bibinfo
  {author} {\bibfnamefont {F.~H.}\ \bibnamefont {da~Jornada}}, \bibinfo
  {author} {\bibfnamefont {Y.}~\bibnamefont {Zhang}}, \bibinfo {author}
  {\bibfnamefont {D.~Y.}\ \bibnamefont {Qiu}}, \bibinfo {author} {\bibfnamefont
  {W.}~\bibnamefont {Ruan}}, \bibinfo {author} {\bibfnamefont {S.-K.}\
  \bibnamefont {Mo}}, \bibinfo {author} {\bibfnamefont {Z.}~\bibnamefont
  {Hussain}}, \bibinfo {author} {\bibfnamefont {Z.-X.}\ \bibnamefont {Shen}},
  \bibinfo {author} {\bibfnamefont {F.}~\bibnamefont {Wang}}, \bibinfo {author}
  {\bibfnamefont {S.~G.}\ \bibnamefont {Louie}},\ and\ \bibinfo {author}
  {\bibfnamefont {M.~F.}\ \bibnamefont {Crommie}},\ }\bibfield  {title}
  {\bibinfo {title} {Giant bandgap renormalization and excitonic effects in a
  monolayer transition metal dichalcogenide semiconductor},\ }\href
  {https://doi.org/10.1038/nmat4061} {\bibfield  {journal} {\bibinfo  {journal}
  {Nat. Mater.}\ }\textbf {\bibinfo {volume} {13}},\ \bibinfo {pages} {1091}
  (\bibinfo {year} {2014})}\BibitemShut {NoStop}%
\bibitem [{\citenamefont {Neaton}\ \emph {et~al.}(2006)\citenamefont {Neaton},
  \citenamefont {Hybertsen},\ and\ \citenamefont {Louie}}]{Neaton2006}%
  \BibitemOpen
  \bibfield  {author} {\bibinfo {author} {\bibfnamefont {J.~B.}\ \bibnamefont
  {Neaton}}, \bibinfo {author} {\bibfnamefont {M.~S.}\ \bibnamefont
  {Hybertsen}},\ and\ \bibinfo {author} {\bibfnamefont {S.~G.}\ \bibnamefont
  {Louie}},\ }\bibfield  {title} {\bibinfo {title} {Renormalization of
  molecular electronic levels at metal-molecule interfaces},\ }\href
  {https://doi.org/10.1103/PhysRevLett.97.216405} {\bibfield  {journal}
  {\bibinfo  {journal} {Phys. Rev. Lett.}\ }\textbf {\bibinfo {volume} {97}},\
  \bibinfo {pages} {216405} (\bibinfo {year} {2006})}\BibitemShut {NoStop}%
\bibitem [{\citenamefont {Thygesen}\ and\ \citenamefont
  {Rubio}(2009)}]{Thygesen2009}%
  \BibitemOpen
  \bibfield  {author} {\bibinfo {author} {\bibfnamefont {K.~S.}\ \bibnamefont
  {Thygesen}}\ and\ \bibinfo {author} {\bibfnamefont {A.}~\bibnamefont
  {Rubio}},\ }\bibfield  {title} {\bibinfo {title} {Renormalization of
  molecular quasiparticle levels at metal-molecule interfaces: Trends across
  binding regimes},\ }\href {https://doi.org/10.1103/PhysRevLett.102.046802}
  {\bibfield  {journal} {\bibinfo  {journal} {Phys. Rev. Lett.}\ }\textbf
  {\bibinfo {volume} {102}},\ \bibinfo {pages} {046802} (\bibinfo {year}
  {2009})}\BibitemShut {NoStop}%
\bibitem [{\citenamefont {Liu}\ \emph {et~al.}(2019)\citenamefont {Liu},
  \citenamefont {da~Jornada}, \citenamefont {Louie},\ and\ \citenamefont
  {Neaton}}]{Liu2019}%
  \BibitemOpen
  \bibfield  {author} {\bibinfo {author} {\bibfnamefont {Z.-F.}\ \bibnamefont
  {Liu}}, \bibinfo {author} {\bibfnamefont {F.~H.}\ \bibnamefont {da~Jornada}},
  \bibinfo {author} {\bibfnamefont {S.~G.}\ \bibnamefont {Louie}},\ and\
  \bibinfo {author} {\bibfnamefont {J.~B.}\ \bibnamefont {Neaton}},\ }\bibfield
   {title} {\bibinfo {title} {Accelerating {GW}-based energy level alignment
  calculations for molecule-metal interfaces using a substrate screening
  approach},\ }\href {https://doi.org/10.1021/acs.jctc.9b00326} {\bibfield
  {journal} {\bibinfo  {journal} {J. Chem. Theory Comput.}\ }\textbf {\bibinfo
  {volume} {15}},\ \bibinfo {pages} {4218} (\bibinfo {year}
  {2019})}\BibitemShut {NoStop}%
\bibitem [{\citenamefont {Xuan}\ \emph {et~al.}(2019)\citenamefont {Xuan},
  \citenamefont {Chen},\ and\ \citenamefont {Quek}}]{Xuan2019}%
  \BibitemOpen
  \bibfield  {author} {\bibinfo {author} {\bibfnamefont {F.}~\bibnamefont
  {Xuan}}, \bibinfo {author} {\bibfnamefont {Y.}~\bibnamefont {Chen}},\ and\
  \bibinfo {author} {\bibfnamefont {S.~Y.}\ \bibnamefont {Quek}},\ }\bibfield
  {title} {\bibinfo {title} {Quasiparticle levels at large interface systems
  from many-body perturbation theory: The {XAF-GW} method},\ }\href
  {https://doi.org/10.1021/acs.jctc.9b00229} {\bibfield  {journal} {\bibinfo
  {journal} {J. Chem. Theory Comput.}\ }\textbf {\bibinfo {volume} {15}},\
  \bibinfo {pages} {3824} (\bibinfo {year} {2019})}\BibitemShut {NoStop}%
\bibitem [{\citenamefont {Frimpong}\ and\ \citenamefont
  {Liu}(2021)}]{Frimpong2021}%
  \BibitemOpen
  \bibfield  {author} {\bibinfo {author} {\bibfnamefont {J.}~\bibnamefont
  {Frimpong}}\ and\ \bibinfo {author} {\bibfnamefont {Z.-F.}\ \bibnamefont
  {Liu}},\ }\bibfield  {title} {\bibinfo {title} {Quasiparticle electronic
  structure of two-dimensional heterotriangulene-based covalent organic
  frameworks adsorbed on {A}u(111)},\ }\href
  {https://doi.org/10.1088/1361-648x/abf7a0} {\bibfield  {journal} {\bibinfo
  {journal} {J. Phys.: Condens. Matter}\ }\textbf {\bibinfo {volume} {33}},\
  \bibinfo {pages} {254004} (\bibinfo {year} {2021})}\BibitemShut {NoStop}%
\bibitem [{\citenamefont {Dr\"{u}ppel}\ \emph {et~al.}(2017)\citenamefont
  {Dr\"{u}ppel}, \citenamefont {Deilmann}, \citenamefont {Kr\"{u}ger},\ and\
  \citenamefont {Rohlfing}}]{Drueppel2017}%
  \BibitemOpen
  \bibfield  {author} {\bibinfo {author} {\bibfnamefont {M.}~\bibnamefont
  {Dr\"{u}ppel}}, \bibinfo {author} {\bibfnamefont {T.}~\bibnamefont
  {Deilmann}}, \bibinfo {author} {\bibfnamefont {P.}~\bibnamefont
  {Kr\"{u}ger}},\ and\ \bibinfo {author} {\bibfnamefont {M.}~\bibnamefont
  {Rohlfing}},\ }\bibfield  {title} {\bibinfo {title} {Diversity of trion
  states and substrate effects in the optical properties of an {M}o{S}$_2$
  monolayer},\ }\href {https://doi.org/10.1038/s41467-017-02286-6} {\bibfield
  {journal} {\bibinfo  {journal} {Nat. Commun.}\ }\textbf {\bibinfo {volume}
  {8}},\ \bibinfo {pages} {2117} (\bibinfo {year} {2017})}\BibitemShut
  {NoStop}%
\bibitem [{\citenamefont {Liu}(2020)}]{Liu2020}%
  \BibitemOpen
  \bibfield  {author} {\bibinfo {author} {\bibfnamefont {Z.-F.}\ \bibnamefont
  {Liu}},\ }\bibfield  {title} {\bibinfo {title} {Dielectric embedding {GW} for
  weakly coupled molecule-metal interfaces},\ }\href
  {https://doi.org/10.1063/1.5140972} {\bibfield  {journal} {\bibinfo
  {journal} {J. Chem. Phys.}\ }\textbf {\bibinfo {volume} {152}},\ \bibinfo
  {pages} {054103} (\bibinfo {year} {2020})}\BibitemShut {NoStop}%
\bibitem [{\citenamefont {Adeniran}\ and\ \citenamefont
  {Liu}(2021)}]{Adeniran2021}%
  \BibitemOpen
  \bibfield  {author} {\bibinfo {author} {\bibfnamefont {O.}~\bibnamefont
  {Adeniran}}\ and\ \bibinfo {author} {\bibfnamefont {Z.-F.}\ \bibnamefont
  {Liu}},\ }\bibfield  {title} {\bibinfo {title} {Quasiparticle electronic
  structure of phthalocyanine:{TMD} interfaces from first-principles {GW}},\
  }\href {https://doi.org/10.1063/5.0072995} {\bibfield  {journal} {\bibinfo
  {journal} {J. Chem. Phys.}\ }\textbf {\bibinfo {volume} {155}},\ \bibinfo
  {pages} {214702} (\bibinfo {year} {2021})}\BibitemShut {NoStop}%
\bibitem [{\citenamefont {Berland}\ and\ \citenamefont
  {Hyldgaard}(2014)}]{Berland2014}%
  \BibitemOpen
  \bibfield  {author} {\bibinfo {author} {\bibfnamefont {K.}~\bibnamefont
  {Berland}}\ and\ \bibinfo {author} {\bibfnamefont {P.}~\bibnamefont
  {Hyldgaard}},\ }\bibfield  {title} {\bibinfo {title} {Exchange functional
  that tests the robustness of the plasmon description of the van der {W}aals
  density functional},\ }\href
  {https://link.aps.org/doi/10.1103/PhysRevB.89.035412} {\bibfield  {journal}
  {\bibinfo  {journal} {Phys. Rev. B}\ }\textbf {\bibinfo {volume} {89}},\
  \bibinfo {pages} {035412} (\bibinfo {year} {2014})}\BibitemShut {NoStop}%
\bibitem [{\citenamefont {Pease}(1952)}]{Pease1952}%
  \BibitemOpen
  \bibfield  {author} {\bibinfo {author} {\bibfnamefont {R.~S.}\ \bibnamefont
  {Pease}},\ }\bibfield  {title} {\bibinfo {title} {An {X}-ray study of boron
  nitride},\ }\href {https://doi.org/10.1107/S0365110X52001064} {\bibfield
  {journal} {\bibinfo  {journal} {Acta Cryst.}\ }\textbf {\bibinfo {volume}
  {5}},\ \bibinfo {pages} {356} (\bibinfo {year} {1952})}\BibitemShut {NoStop}%
\bibitem [{\citenamefont {Wickramaratne}\ \emph {et~al.}(2018)\citenamefont
  {Wickramaratne}, \citenamefont {Weston},\ and\ \citenamefont {Van~de
  Walle}}]{Wickramaratne2018}%
  \BibitemOpen
  \bibfield  {author} {\bibinfo {author} {\bibfnamefont {D.}~\bibnamefont
  {Wickramaratne}}, \bibinfo {author} {\bibfnamefont {L.}~\bibnamefont
  {Weston}},\ and\ \bibinfo {author} {\bibfnamefont {C.~G.}\ \bibnamefont
  {Van~de Walle}},\ }\bibfield  {title} {\bibinfo {title} {Monolayer to bulk
  properties of hexagonal boron nitride},\ }\href
  {https://doi.org/10.1021/acs.jpcc.8b09087} {\bibfield  {journal} {\bibinfo
  {journal} {J. Phys. Chem. C}\ }\textbf {\bibinfo {volume} {122}},\ \bibinfo
  {pages} {25524} (\bibinfo {year} {2018})}\BibitemShut {NoStop}%
\bibitem [{\citenamefont {Schlipf}\ and\ \citenamefont
  {Gygi}(2015)}]{Schlipf2015}%
  \BibitemOpen
  \bibfield  {author} {\bibinfo {author} {\bibfnamefont {M.}~\bibnamefont
  {Schlipf}}\ and\ \bibinfo {author} {\bibfnamefont {F.}~\bibnamefont {Gygi}},\
  }\bibfield  {title} {\bibinfo {title} {Optimization algorithm for the
  generation of {ONCV} pseudopotentials},\ }\href
  {https://doi.org/https://doi.org/10.1016/j.cpc.2015.05.011} {\bibfield
  {journal} {\bibinfo  {journal} {Comput. Phys. Commun.}\ }\textbf {\bibinfo
  {volume} {196}},\ \bibinfo {pages} {36} (\bibinfo {year} {2015})}\BibitemShut
  {NoStop}%
\bibitem [{\citenamefont {Hamann}(2013)}]{Hamann2013}%
  \BibitemOpen
  \bibfield  {author} {\bibinfo {author} {\bibfnamefont {D.~R.}\ \bibnamefont
  {Hamann}},\ }\bibfield  {title} {\bibinfo {title} {Optimized norm-conserving
  {V}anderbilt pseudopotentials},\ }\href
  {https://doi.org/10.1103/PhysRevB.88.085117} {\bibfield  {journal} {\bibinfo
  {journal} {Phys. Rev. B}\ }\textbf {\bibinfo {volume} {88}},\ \bibinfo
  {pages} {085117} (\bibinfo {year} {2013})}\BibitemShut {NoStop}%
\bibitem [{\citenamefont {Giannozzi}\ \emph {et~al.}(2017)\citenamefont
  {Giannozzi}, \citenamefont {Andreussi}, \citenamefont {Brumme}, \citenamefont
  {Bunau}, \citenamefont {Buongiorno~Nardelli}, \citenamefont {Calandra},
  \citenamefont {Car}, \citenamefont {Cavazzoni}, \citenamefont {Ceresoli},
  \citenamefont {Cococcioni}, \citenamefont {Colonna}, \citenamefont
  {Carnimeo}, \citenamefont {Dal~Corso}, \citenamefont {de~Gironcoli},
  \citenamefont {Delugas}, \citenamefont {DiStasio}, \citenamefont {Ferretti},
  \citenamefont {Floris}, \citenamefont {Fratesi}, \citenamefont {Fugallo},
  \citenamefont {Gebauer}, \citenamefont {Gerstmann}, \citenamefont {Giustino},
  \citenamefont {Gorni}, \citenamefont {Jia}, \citenamefont {Kawamura},
  \citenamefont {Ko}, \citenamefont {Kokalj}, \citenamefont
  {K\"{u}\c{c}\"{u}kbenli}, \citenamefont {Lazzeri}, \citenamefont {Marsili},
  \citenamefont {Marzari}, \citenamefont {Mauri}, \citenamefont {Nguyen},
  \citenamefont {Nguyen}, \citenamefont {Otero-de-la Roza}, \citenamefont
  {Paulatto}, \citenamefont {Ponc\'{e}}, \citenamefont {Rocca}, \citenamefont
  {Sabatini}, \citenamefont {Santra}, \citenamefont {Schlipf}, \citenamefont
  {Seitsonen}, \citenamefont {Smogunov}, \citenamefont {Timrov}, \citenamefont
  {Thonhauser}, \citenamefont {Umari}, \citenamefont {Vast}, \citenamefont
  {Wu},\ and\ \citenamefont {Baroni}}]{Giannozzi2017}%
  \BibitemOpen
  \bibfield  {author} {\bibinfo {author} {\bibfnamefont {P.}~\bibnamefont
  {Giannozzi}}, \bibinfo {author} {\bibfnamefont {O.}~\bibnamefont
  {Andreussi}}, \bibinfo {author} {\bibfnamefont {T.}~\bibnamefont {Brumme}},
  \bibinfo {author} {\bibfnamefont {O.}~\bibnamefont {Bunau}}, \bibinfo
  {author} {\bibfnamefont {M.}~\bibnamefont {Buongiorno~Nardelli}}, \bibinfo
  {author} {\bibfnamefont {M.}~\bibnamefont {Calandra}}, \bibinfo {author}
  {\bibfnamefont {R.}~\bibnamefont {Car}}, \bibinfo {author} {\bibfnamefont
  {C.}~\bibnamefont {Cavazzoni}}, \bibinfo {author} {\bibfnamefont
  {D.}~\bibnamefont {Ceresoli}}, \bibinfo {author} {\bibfnamefont
  {M.}~\bibnamefont {Cococcioni}}, \bibinfo {author} {\bibfnamefont
  {N.}~\bibnamefont {Colonna}}, \bibinfo {author} {\bibfnamefont
  {I.}~\bibnamefont {Carnimeo}}, \bibinfo {author} {\bibfnamefont
  {A.}~\bibnamefont {Dal~Corso}}, \bibinfo {author} {\bibfnamefont
  {S.}~\bibnamefont {de~Gironcoli}}, \bibinfo {author} {\bibfnamefont
  {P.}~\bibnamefont {Delugas}}, \bibinfo {author} {\bibfnamefont {R.~A.}\
  \bibnamefont {DiStasio}}, \bibinfo {author} {\bibfnamefont {A.}~\bibnamefont
  {Ferretti}}, \bibinfo {author} {\bibfnamefont {A.}~\bibnamefont {Floris}},
  \bibinfo {author} {\bibfnamefont {G.}~\bibnamefont {Fratesi}}, \bibinfo
  {author} {\bibfnamefont {G.}~\bibnamefont {Fugallo}}, \bibinfo {author}
  {\bibfnamefont {R.}~\bibnamefont {Gebauer}}, \bibinfo {author} {\bibfnamefont
  {U.}~\bibnamefont {Gerstmann}}, \bibinfo {author} {\bibfnamefont
  {F.}~\bibnamefont {Giustino}}, \bibinfo {author} {\bibfnamefont
  {T.}~\bibnamefont {Gorni}}, \bibinfo {author} {\bibfnamefont
  {J.}~\bibnamefont {Jia}}, \bibinfo {author} {\bibfnamefont {M.}~\bibnamefont
  {Kawamura}}, \bibinfo {author} {\bibfnamefont {H.-Y.}\ \bibnamefont {Ko}},
  \bibinfo {author} {\bibfnamefont {A.}~\bibnamefont {Kokalj}}, \bibinfo
  {author} {\bibfnamefont {E.}~\bibnamefont {K\"{u}\c{c}\"{u}kbenli}}, \bibinfo
  {author} {\bibfnamefont {M.}~\bibnamefont {Lazzeri}}, \bibinfo {author}
  {\bibfnamefont {M.}~\bibnamefont {Marsili}}, \bibinfo {author} {\bibfnamefont
  {N.}~\bibnamefont {Marzari}}, \bibinfo {author} {\bibfnamefont
  {F.}~\bibnamefont {Mauri}}, \bibinfo {author} {\bibfnamefont {N.~L.}\
  \bibnamefont {Nguyen}}, \bibinfo {author} {\bibfnamefont {H.-V.}\
  \bibnamefont {Nguyen}}, \bibinfo {author} {\bibfnamefont {A.}~\bibnamefont
  {Otero-de-la Roza}}, \bibinfo {author} {\bibfnamefont {L.}~\bibnamefont
  {Paulatto}}, \bibinfo {author} {\bibfnamefont {S.}~\bibnamefont {Ponc\'{e}}},
  \bibinfo {author} {\bibfnamefont {D.}~\bibnamefont {Rocca}}, \bibinfo
  {author} {\bibfnamefont {R.}~\bibnamefont {Sabatini}}, \bibinfo {author}
  {\bibfnamefont {B.}~\bibnamefont {Santra}}, \bibinfo {author} {\bibfnamefont
  {M.}~\bibnamefont {Schlipf}}, \bibinfo {author} {\bibfnamefont {A.~P.}\
  \bibnamefont {Seitsonen}}, \bibinfo {author} {\bibfnamefont {A.}~\bibnamefont
  {Smogunov}}, \bibinfo {author} {\bibfnamefont {I.}~\bibnamefont {Timrov}},
  \bibinfo {author} {\bibfnamefont {T.}~\bibnamefont {Thonhauser}}, \bibinfo
  {author} {\bibfnamefont {P.}~\bibnamefont {Umari}}, \bibinfo {author}
  {\bibfnamefont {N.}~\bibnamefont {Vast}}, \bibinfo {author} {\bibfnamefont
  {X.}~\bibnamefont {Wu}},\ and\ \bibinfo {author} {\bibfnamefont
  {S.}~\bibnamefont {Baroni}},\ }\bibfield  {title} {\bibinfo {title} {Advanced
  capabilities for materials modelling with {Q}uantum {ESPRESSO}},\ }\href
  {https://doi.org/10.1088/1361-648x/aa8f79} {\bibfield  {journal} {\bibinfo
  {journal} {J. Phys.: Condens. Matter}\ }\textbf {\bibinfo {volume} {29}},\
  \bibinfo {pages} {465901} (\bibinfo {year} {2017})}\BibitemShut {NoStop}%
\bibitem [{\citenamefont {Momma}\ and\ \citenamefont
  {Izumi}(2008)}]{Momma2008}%
  \BibitemOpen
  \bibfield  {author} {\bibinfo {author} {\bibfnamefont {K.}~\bibnamefont
  {Momma}}\ and\ \bibinfo {author} {\bibfnamefont {F.}~\bibnamefont {Izumi}},\
  }\bibfield  {title} {\bibinfo {title} {{VESTA}: a three-dimensional
  visualization system for electronic and structural analysis},\ }\href
  {https://doi.org/10.1107/s0021889808012016} {\bibfield  {journal} {\bibinfo
  {journal} {J. Appl. Cryst.}\ }\textbf {\bibinfo {volume} {41}},\ \bibinfo
  {pages} {653} (\bibinfo {year} {2008})}\BibitemShut {NoStop}%
\bibitem [{\citenamefont {{Deslippe}}\ \emph {et~al.}(2012)\citenamefont
  {{Deslippe}}, \citenamefont {{Samsonidze}}, \citenamefont {{Strubbe}},
  \citenamefont {{Jain}}, \citenamefont {{Cohen}},\ and\ \citenamefont
  {{Louie}}}]{Deslippe2012}%
  \BibitemOpen
  \bibfield  {author} {\bibinfo {author} {\bibfnamefont {J.}~\bibnamefont
  {{Deslippe}}}, \bibinfo {author} {\bibfnamefont {G.}~\bibnamefont
  {{Samsonidze}}}, \bibinfo {author} {\bibfnamefont {D.~A.}\ \bibnamefont
  {{Strubbe}}}, \bibinfo {author} {\bibfnamefont {M.}~\bibnamefont {{Jain}}},
  \bibinfo {author} {\bibfnamefont {M.~L.}\ \bibnamefont {{Cohen}}},\ and\
  \bibinfo {author} {\bibfnamefont {S.~G.}\ \bibnamefont {{Louie}}},\
  }\bibfield  {title} {\bibinfo {title} {{BerkeleyGW: A massively parallel
  computer package for the calculation of the quasiparticle and optical
  properties of materials and nanostructures}},\ }\href
  {https://doi.org/10.1016/j.cpc.2011.12.006} {\bibfield  {journal} {\bibinfo
  {journal} {Comput. Phys. Commun.}\ }\textbf {\bibinfo {volume} {183}},\
  \bibinfo {pages} {1269} (\bibinfo {year} {2012})}\BibitemShut {NoStop}%
\bibitem [{\citenamefont {Perdew}\ \emph {et~al.}(1996)\citenamefont {Perdew},
  \citenamefont {Burke},\ and\ \citenamefont {Ernzerhof}}]{Perdew1996}%
  \BibitemOpen
  \bibfield  {author} {\bibinfo {author} {\bibfnamefont {J.~P.}\ \bibnamefont
  {Perdew}}, \bibinfo {author} {\bibfnamefont {K.}~\bibnamefont {Burke}},\ and\
  \bibinfo {author} {\bibfnamefont {M.}~\bibnamefont {Ernzerhof}},\ }\bibfield
  {title} {\bibinfo {title} {Generalized gradient approximation made simple},\
  }\href {https://doi.org/10.1103/PhysRevLett.77.3865} {\bibfield  {journal}
  {\bibinfo  {journal} {Phys. Rev. Lett.}\ }\textbf {\bibinfo {volume} {77}},\
  \bibinfo {pages} {3865} (\bibinfo {year} {1996})}\BibitemShut {NoStop}%
\bibitem [{\citenamefont {Ismail-Beigi}(2006)}]{IsmailBeigi2006}%
  \BibitemOpen
  \bibfield  {author} {\bibinfo {author} {\bibfnamefont {S.}~\bibnamefont
  {Ismail-Beigi}},\ }\bibfield  {title} {\bibinfo {title} {Truncation of
  periodic image interactions for confined systems},\ }\href
  {https://doi.org/10.1103/PhysRevB.73.233103} {\bibfield  {journal} {\bibinfo
  {journal} {Phys. Rev. B}\ }\textbf {\bibinfo {volume} {73}},\ \bibinfo
  {pages} {233103} (\bibinfo {year} {2006})}\BibitemShut {NoStop}%
\bibitem [{\citenamefont {Deslippe}\ \emph {et~al.}(2013)\citenamefont
  {Deslippe}, \citenamefont {Samsonidze}, \citenamefont {Jain}, \citenamefont
  {Cohen},\ and\ \citenamefont {Louie}}]{Deslippe2013}%
  \BibitemOpen
  \bibfield  {author} {\bibinfo {author} {\bibfnamefont {J.}~\bibnamefont
  {Deslippe}}, \bibinfo {author} {\bibfnamefont {G.}~\bibnamefont
  {Samsonidze}}, \bibinfo {author} {\bibfnamefont {M.}~\bibnamefont {Jain}},
  \bibinfo {author} {\bibfnamefont {M.~L.}\ \bibnamefont {Cohen}},\ and\
  \bibinfo {author} {\bibfnamefont {S.~G.}\ \bibnamefont {Louie}},\ }\bibfield
  {title} {\bibinfo {title} {Coulomb-hole summations and energies for {GW}
  calculations with limited number of empty orbitals: A modified static
  remainder approach},\ }\href
  {https://link.aps.org/doi/10.1103/PhysRevB.87.165124} {\bibfield  {journal}
  {\bibinfo  {journal} {Phys. Rev. B}\ }\textbf {\bibinfo {volume} {87}},\
  \bibinfo {pages} {165124} (\bibinfo {year} {2013})}\BibitemShut {NoStop}%
\bibitem [{\citenamefont {Trolle}\ \emph {et~al.}(2017)\citenamefont {Trolle},
  \citenamefont {Pedersen},\ and\ \citenamefont {V\'{e}niard}}]{Trolle2017}%
  \BibitemOpen
  \bibfield  {author} {\bibinfo {author} {\bibfnamefont {M.~L.}\ \bibnamefont
  {Trolle}}, \bibinfo {author} {\bibfnamefont {T.~G.}\ \bibnamefont
  {Pedersen}},\ and\ \bibinfo {author} {\bibfnamefont {V.}~\bibnamefont
  {V\'{e}niard}},\ }\bibfield  {title} {\bibinfo {title} {Model dielectric
  function for 2{D} semiconductors including substrate screening},\ }\href@noop
  {} {\bibfield  {journal} {\bibinfo  {journal} {Sci. Rep.}\ }\textbf {\bibinfo
  {volume} {7}},\ \bibinfo {pages} {39844} (\bibinfo {year}
  {2017})}\BibitemShut {NoStop}%
\bibitem [{\citenamefont {Br\"{u}lke}\ \emph {et~al.}(2020)\citenamefont
  {Br\"{u}lke}, \citenamefont {Bauer},\ and\ \citenamefont
  {Sokolowski}}]{Bruelke2020}%
  \BibitemOpen
  \bibfield  {author} {\bibinfo {author} {\bibfnamefont {C.}~\bibnamefont
  {Br\"{u}lke}}, \bibinfo {author} {\bibfnamefont {O.}~\bibnamefont {Bauer}},\
  and\ \bibinfo {author} {\bibfnamefont {M.~M.}\ \bibnamefont {Sokolowski}},\
  }\bibfield  {title} {\bibinfo {title} {The influence of an interfacial h{BN}
  layer on the fluorescence of an organic molecule},\ }\href@noop {} {\bibfield
   {journal} {\bibinfo  {journal} {Beilstein J Nanotechnol.}\ }\textbf
  {\bibinfo {volume} {11}},\ \bibinfo {pages} {1663} (\bibinfo {year}
  {2020})}\BibitemShut {NoStop}%
\bibitem [{\citenamefont {Mak}\ \emph {et~al.}(2010)\citenamefont {Mak},
  \citenamefont {Lee}, \citenamefont {Hone}, \citenamefont {Shan},\ and\
  \citenamefont {Heinz}}]{Mak2010}%
  \BibitemOpen
  \bibfield  {author} {\bibinfo {author} {\bibfnamefont {K.~F.}\ \bibnamefont
  {Mak}}, \bibinfo {author} {\bibfnamefont {C.}~\bibnamefont {Lee}}, \bibinfo
  {author} {\bibfnamefont {J.}~\bibnamefont {Hone}}, \bibinfo {author}
  {\bibfnamefont {J.}~\bibnamefont {Shan}},\ and\ \bibinfo {author}
  {\bibfnamefont {T.~F.}\ \bibnamefont {Heinz}},\ }\bibfield  {title} {\bibinfo
  {title} {Atomically thin {M}o{S}$_2$: {A} new direct-gap semiconductor},\
  }\href@noop {} {\bibfield  {journal} {\bibinfo  {journal} {Phys. Rev. Lett.}\
  }\textbf {\bibinfo {volume} {105}},\ \bibinfo {pages} {136805} (\bibinfo
  {year} {2010})}\BibitemShut {NoStop}%
\bibitem [{\citenamefont {Utama}\ \emph {et~al.}(2019)\citenamefont {Utama},
  \citenamefont {Kleemann}, \citenamefont {Zhao}, \citenamefont {Ong},
  \citenamefont {da~Jornada}, \citenamefont {Qiu}, \citenamefont {Cai},
  \citenamefont {Li}, \citenamefont {Kou}, \citenamefont {Zhao}, \citenamefont
  {Wang}, \citenamefont {Watanabe}, \citenamefont {Taniguchi}, \citenamefont
  {Tongay}, \citenamefont {Zettl}, \citenamefont {Louie},\ and\ \citenamefont
  {Wang}}]{Utama2019}%
  \BibitemOpen
  \bibfield  {author} {\bibinfo {author} {\bibfnamefont {M.~I.~B.}\
  \bibnamefont {Utama}}, \bibinfo {author} {\bibfnamefont {H.}~\bibnamefont
  {Kleemann}}, \bibinfo {author} {\bibfnamefont {W.}~\bibnamefont {Zhao}},
  \bibinfo {author} {\bibfnamefont {C.~S.}\ \bibnamefont {Ong}}, \bibinfo
  {author} {\bibfnamefont {F.~H.}\ \bibnamefont {da~Jornada}}, \bibinfo
  {author} {\bibfnamefont {D.~Y.}\ \bibnamefont {Qiu}}, \bibinfo {author}
  {\bibfnamefont {H.}~\bibnamefont {Cai}}, \bibinfo {author} {\bibfnamefont
  {H.}~\bibnamefont {Li}}, \bibinfo {author} {\bibfnamefont {R.}~\bibnamefont
  {Kou}}, \bibinfo {author} {\bibfnamefont {S.}~\bibnamefont {Zhao}}, \bibinfo
  {author} {\bibfnamefont {S.}~\bibnamefont {Wang}}, \bibinfo {author}
  {\bibfnamefont {K.}~\bibnamefont {Watanabe}}, \bibinfo {author}
  {\bibfnamefont {T.}~\bibnamefont {Taniguchi}}, \bibinfo {author}
  {\bibfnamefont {S.}~\bibnamefont {Tongay}}, \bibinfo {author} {\bibfnamefont
  {A.}~\bibnamefont {Zettl}}, \bibinfo {author} {\bibfnamefont {S.~G.}\
  \bibnamefont {Louie}},\ and\ \bibinfo {author} {\bibfnamefont
  {F.}~\bibnamefont {Wang}},\ }\bibfield  {title} {\bibinfo {title} {A
  dielectric-defined lateral heterojunction in a monolayer semiconductor},\
  }\href {https://doi.org/10.1038/s41928-019-0207-4} {\bibfield  {journal}
  {\bibinfo  {journal} {Nat. Electron.}\ }\textbf {\bibinfo {volume} {2}},\
  \bibinfo {pages} {60} (\bibinfo {year} {2019})}\BibitemShut {NoStop}%
\bibitem [{\citenamefont {Rohlfing}(2010)}]{Rohlfing2010}%
  \BibitemOpen
  \bibfield  {author} {\bibinfo {author} {\bibfnamefont {M.}~\bibnamefont
  {Rohlfing}},\ }\bibfield  {title} {\bibinfo {title} {Electronic excitations
  from a perturbative {LDA}+{$GdW$} approach},\ }\href@noop {} {\bibfield
  {journal} {\bibinfo  {journal} {Phys. Rev. B}\ }\textbf {\bibinfo {volume}
  {82}},\ \bibinfo {pages} {205127} (\bibinfo {year} {2010})}\BibitemShut
  {NoStop}%
\bibitem [{\citenamefont {Naik}\ and\ \citenamefont {Jain}(2018)}]{Naik2018}%
  \BibitemOpen
  \bibfield  {author} {\bibinfo {author} {\bibfnamefont {M.~H.}\ \bibnamefont
  {Naik}}\ and\ \bibinfo {author} {\bibfnamefont {M.}~\bibnamefont {Jain}},\
  }\bibfield  {title} {\bibinfo {title} {Substrate screening effects on the
  quasiparticle band gap and defect charge transition levels in {M}o{S}$_2$},\
  }\href {https://doi.org/10.1103/PhysRevMaterials.2.084002} {\bibfield
  {journal} {\bibinfo  {journal} {Phys. Rev. Mater.}\ }\textbf {\bibinfo
  {volume} {2}},\ \bibinfo {pages} {084002} (\bibinfo {year}
  {2018})}\BibitemShut {NoStop}%
\bibitem [{\citenamefont {Rizzo}\ \emph {et~al.}(2020)\citenamefont {Rizzo},
  \citenamefont {Dai}, \citenamefont {Bronner}, \citenamefont {Veber},
  \citenamefont {Smith}, \citenamefont {Matsumoto}, \citenamefont {Thomas},
  \citenamefont {Nguyen}, \citenamefont {Forrester}, \citenamefont {Zhao},
  \citenamefont {J\o{}rgensen}, \citenamefont {Dichtel}, \citenamefont
  {Fischer}, \citenamefont {Li}, \citenamefont {Bredas},\ and\ \citenamefont
  {Crommie}}]{Rizzo2020}%
  \BibitemOpen
  \bibfield  {author} {\bibinfo {author} {\bibfnamefont {D.~J.}\ \bibnamefont
  {Rizzo}}, \bibinfo {author} {\bibfnamefont {Q.}~\bibnamefont {Dai}}, \bibinfo
  {author} {\bibfnamefont {C.}~\bibnamefont {Bronner}}, \bibinfo {author}
  {\bibfnamefont {G.}~\bibnamefont {Veber}}, \bibinfo {author} {\bibfnamefont
  {B.~J.}\ \bibnamefont {Smith}}, \bibinfo {author} {\bibfnamefont
  {M.}~\bibnamefont {Matsumoto}}, \bibinfo {author} {\bibfnamefont
  {S.}~\bibnamefont {Thomas}}, \bibinfo {author} {\bibfnamefont {G.~D.}\
  \bibnamefont {Nguyen}}, \bibinfo {author} {\bibfnamefont {P.~R.}\
  \bibnamefont {Forrester}}, \bibinfo {author} {\bibfnamefont {W.}~\bibnamefont
  {Zhao}}, \bibinfo {author} {\bibfnamefont {J.~H.}\ \bibnamefont
  {J\o{}rgensen}}, \bibinfo {author} {\bibfnamefont {W.~R.}\ \bibnamefont
  {Dichtel}}, \bibinfo {author} {\bibfnamefont {F.~R.}\ \bibnamefont
  {Fischer}}, \bibinfo {author} {\bibfnamefont {H.}~\bibnamefont {Li}},
  \bibinfo {author} {\bibfnamefont {J.-L.}\ \bibnamefont {Bredas}},\ and\
  \bibinfo {author} {\bibfnamefont {M.~F.}\ \bibnamefont {Crommie}},\
  }\bibfield  {title} {\bibinfo {title} {Revealing the local electronic
  structure of a single-layer covalent organic framework through electronic
  decoupling},\ }\href@noop {} {\bibfield  {journal} {\bibinfo  {journal} {Nano
  Lett.}\ }\textbf {\bibinfo {volume} {20}},\ \bibinfo {pages} {963} (\bibinfo
  {year} {2020})}\BibitemShut {NoStop}%
\bibitem [{\citenamefont {Debbichi}\ \emph {et~al.}(2014)\citenamefont
  {Debbichi}, \citenamefont {Eriksson},\ and\ \citenamefont
  {Leb\`{e}gue}}]{Debbichi2014}%
  \BibitemOpen
  \bibfield  {author} {\bibinfo {author} {\bibfnamefont {L.}~\bibnamefont
  {Debbichi}}, \bibinfo {author} {\bibfnamefont {O.}~\bibnamefont {Eriksson}},\
  and\ \bibinfo {author} {\bibfnamefont {S.}~\bibnamefont {Leb\`{e}gue}},\
  }\bibfield  {title} {\bibinfo {title} {Electronic structure of
  two-dimensional transition metal dichalcogenide bilayers from ab initio
  theory},\ }\href {https://doi.org/10.1103/PhysRevB.89.205311} {\bibfield
  {journal} {\bibinfo  {journal} {Phys. Rev. B}\ }\textbf {\bibinfo {volume}
  {89}},\ \bibinfo {pages} {205311} (\bibinfo {year} {2014})}\BibitemShut
  {NoStop}%
\bibitem [{\citenamefont {Gao}\ \emph {et~al.}(2017)\citenamefont {Gao},
  \citenamefont {Yang},\ and\ \citenamefont {Spataru}}]{Gao2017}%
  \BibitemOpen
  \bibfield  {author} {\bibinfo {author} {\bibfnamefont {S.}~\bibnamefont
  {Gao}}, \bibinfo {author} {\bibfnamefont {L.}~\bibnamefont {Yang}},\ and\
  \bibinfo {author} {\bibfnamefont {C.~D.}\ \bibnamefont {Spataru}},\
  }\bibfield  {title} {\bibinfo {title} {Interlayer coupling and gate-tunable
  excitons in transition metal dichalcogenide heterostructures},\ }\href@noop
  {} {\bibfield  {journal} {\bibinfo  {journal} {Nano Lett.}\ }\textbf
  {\bibinfo {volume} {17}},\ \bibinfo {pages} {7809} (\bibinfo {year}
  {2017})}\BibitemShut {NoStop}%
\bibitem [{\citenamefont {Laturia}\ \emph {et~al.}(2018)\citenamefont
  {Laturia}, \citenamefont {Van~de Put},\ and\ \citenamefont
  {Vandenberghe}}]{Laturia2018}%
  \BibitemOpen
  \bibfield  {author} {\bibinfo {author} {\bibfnamefont {A.}~\bibnamefont
  {Laturia}}, \bibinfo {author} {\bibfnamefont {M.~L.}\ \bibnamefont {Van~de
  Put}},\ and\ \bibinfo {author} {\bibfnamefont {W.~G.}\ \bibnamefont
  {Vandenberghe}},\ }\bibfield  {title} {\bibinfo {title} {Dielectric
  properties of hexagonal boron nitride and transition metal dichalcogenides:
  from monolayer to bulk},\ }\href@noop {} {\bibfield  {journal} {\bibinfo
  {journal} {npj 2D Mater. Appl.}\ }\textbf {\bibinfo {volume} {2}},\ \bibinfo
  {pages} {6} (\bibinfo {year} {2018})}\BibitemShut {NoStop}%
\bibitem [{\citenamefont {Bernardi}\ \emph {et~al.}(2013)\citenamefont
  {Bernardi}, \citenamefont {Palummo},\ and\ \citenamefont
  {Grossman}}]{Bernardi2013}%
  \BibitemOpen
  \bibfield  {author} {\bibinfo {author} {\bibfnamefont {M.}~\bibnamefont
  {Bernardi}}, \bibinfo {author} {\bibfnamefont {M.}~\bibnamefont {Palummo}},\
  and\ \bibinfo {author} {\bibfnamefont {J.~C.}\ \bibnamefont {Grossman}},\
  }\bibfield  {title} {\bibinfo {title} {Extraordinary sunlight absorption and
  one nanometer thick photovoltaics using two-dimensional monolayer
  materials},\ }\href {https://doi.org/10.1021/nl401544y} {\bibfield  {journal}
  {\bibinfo  {journal} {Nano Lett.}\ }\textbf {\bibinfo {volume} {13}},\
  \bibinfo {pages} {3664} (\bibinfo {year} {2013})}\BibitemShut {NoStop}%
\end{thebibliography}%
\end{document}